\documentclass{aastex62}

\usepackage{amsmath}
\usepackage{amssymb}
\usepackage{subfig}
\usepackage{placeins}

\received{April 23, 2016}
\revised{July 9, 2016}
\accepted{July 14, 2016}
\published{September 23, 2016}
\submitjournal{The Astrophysical Journal}

\shorttitle{Damping of Nonlinear Density Waves in Dense Planetary Rings}
\shortauthors{Lehmann et al.}

\begin{document}

\title{A WEAKLY NONLINEAR MODEL FOR THE DAMPING OF RESONANTLY FORCED DENSITY WAVES IN DENSE PLANETARY RINGS}

\correspondingauthor{Marius Lehmann}
\email{marius.lehmann@oulu.fi}

\author[0000-0002-0496-3539]{Marius Lehmann}
\affil{Astronomy Research Unit, University of Oulu, Finland}

\author{J\"urgen Schmidt}
\affil{Astronomy Research Unit, University of Oulu, Finland}

\author[0000-0002-4400-042X]{Heikki Salo}
\affil{Astronomy Research Unit, University of Oulu, Finland}

%
%
%
%



\begin{abstract}

In this paper we address the stability of resonantly forced density waves in dense planetary rings.
  Already by \citet{goldreich1978b} it has been argued that density waves might be unstable, depending on the relationship between the ring's viscosity and the 
surface mass density.
  In the recent paper \citet{schmidt2016} we have pointed out that when - within a fluid description of the ring dynamics - the criterion for viscous 
overstability is satisfied, forced spiral density waves become unstable as well.
  In this case, linear theory fails to describe the damping, but nonlinearity of the underlying equations guarantees a finite amplitude and eventually a damping 
of the wave. 
  We apply the multiple scale formalism to derive a weakly nonlinear damping relation from a hydrodynamical model. 
  This relation describes the resonant excitation and nonlinear viscous damping of spiral density waves in a vertically integrated fluid disk with density 
dependent transport coefficients.
  The model consistently predicts density waves to be (linearly) unstable in a ring region where the conditions for viscous overstability are met.
  Sufficiently far away from the Lindblad resonance, the surface mass density perturbation is predicted to saturate to a constant value due to nonlinear viscous 
damping.
  The wave's damping lengths of the model depend on certain input parameters, such as the distance to the threshold for viscous overstability in parameter space 
and the ground state surface mass density.

\end{abstract}



\section{Introduction}\label{sec:intro}
Density wave theory, originally proposed to explain the spiral structure of galactic disks, has been applied
to wave trains excited at resonances in Saturn's rings, notably the inner Lindblad resonances (ILR) of various Saturnian
satellites.

The linear fluid description of density waves is well developed (\citet{goldreich1979c}; \citet{shu1984}) and it has been applied to derive estimates on the 
local surface mass density of Saturn's rings (\citet{esposito1983c};~\citet{lissauer1985}; 
~\citet{nicholson1990a};~\citet{rosen1991a};~\citet{rosen1991b};~\citet{2004Icar..171..372S};~\citet{tiscareno2007b}).
Also, estimates for the ring viscosity have been obtained from the observed damping of density waves. These are based on a linear damping relation with the 
assumption of a constant viscosity.  
However, already \citet{goldreich1978b} pointed out that linear density waves might also be unstable, depending on the behavior of the viscosity as a function 
of the local surface mass density of the rings.
If a density wave is unstable, a linear description is not applicable.

Traditionally, the theory of the excitation and damping of ``nonlinear density waves'' in Saturn's rings is based on a description of the ring matter as 
Lagrangian fluid elements (streamlines) that interact with each other in terms of perturbations of their orbital elements (\citet{Borderies83}, 
\citet{borderies1983c}, \citet{borderies1985}, \citet{shu1985a}, \citet{shu1985b}, \citet{longaretti1986}, \citet{bgt1986}, \citet{longaretti1987}, 
\citet{borderies1989}). 
In this formalism the degree of nonlinearity (of a wave) is a direct measure of the relative spacing between adjacent streamlines.
\citet{borderies1985} applied this streamline formalism to describe the propagation and damping of spiral density waves as self-excited features in fluid disks. 
They computed the pressure tensor from 
a fluid model describing a collection of densely packed inelastic spheres (\citet{haff1983b}). The pressure tensor determines the damping behavior of the waves 
and they concluded that (free) density waves should be unstable in closely packed rings and stable 
in dilute rings. 
\citet{bgt1986} derived radial profiles of tightly wound nonlinear density waves by computing for different viscosity models the isothermal steady state 
structure of a ring patch in the presence of a density wave.
They used a damping relation for waves derived from the conservation of angular momentum luminosity and the nonlinear dispersion relation for tightly wound 
density waves, as derived with the streamline formalism.
\citet{shu1985a} and \citet{shu1985b} adopted a similar approach. They also took into account the local heating of the ring due to the density wave itself in 
their calculation of the pressure tensor from the second-order moment of the Boltzmann equation.
They used a Krook ansatz to account for the effect of particle collisions.
\citet{longaretti1987} applied the streamline formalism to a profile of the Mimas 5:3 density wave, obtained by the Voyager Photo Polarimeter Subsystem, to 
retrieve
the radial dependences of all the functions involved in this formalism. The resulting theoretical surface density profile fits the Voyager data quite well.
Later, \citet{rappaport2009} applied an improved inversion procedure to derive essentially the same functions, but with higher accuracy, since their procedure 
is based on a 
combination of multiple optical depth profiles obtained from data from the Cassini Radio Science Subsystem (RSS).
Conclusions of \citet{bgt1986} and \citet{shu1985b} were that their nonlinear density wave models can successfully describe many features of observed strong 
density waves in Saturn's rings, but also that they underestimate the wave damping in dense ring regions.

Recently, the applicability of the \emph{linear} hydrodynamic damping relation for density waves has been revisited by \citet{schmidt2016} in relation to the 
strong variety of damping lengths among the first order Janus waves, as seen in Cassini data. 
In that paper it has been shown that, in a fluid description of dense rings, a linear density wave is unstable if the criterion for the onset of viscous 
overstability is fulfilled.
The viscous overstability is an intrinsic instability of the ring flow which leads to the formation of axi-symmetric wave trains with typical wavelengths 
$\lambda \sim 100\text{m}-200\text{m}$ 
(\citet{schmit1995};~\citet{schmit1999};~\citet{schmidt2001b};~\citet{salo2001};~\citet{schmidt2003};~\citet{latter2009};~\citet{latter2010}). The occurrence of 
this periodic micro structure in the dense Saturnian rings has been confirmed in Cassini RSS and Ultra Violet Imaging Spectrograph (UVIS) data 
(\citet{Thomson2007};~(\citet{colwell2007})).
This leads to the conclusion that in some regions of Saturn's rings density waves might suffer from linear instability and one would expect that these waves 
show a qualitatively different damping behavior than the one
predicted by linear theory, even for weakly forced waves.

It is the main goal of this paper to approach the problem of an unstable density wave according to the scenario pictured above in terms of a fluid model. The 
wave pattern, which forms
at resonance is also responsive to the linear instability of the ring flow. The latter is controlled by a threshold parameter, which is in this case the density 
dependence of the viscosity,
a criterion directly related to the viscous overstability. Not too far away from the instability threshold, one expects that
the wave pattern obeys a cubic amplitude equation (\citet{cross1993}). By performing a multiple scale expansion (\citet{kevorkian1996}; \citet{cross1993}) of 
the hydrodynamical equations we derive a weakly nonlinear model,
which describes the damping of resonantly forced spiral density waves in a vertically integrated fluid disk with density dependent transport coefficients in 
terms of such a cubic amplitude equation.

In the idealized model, a density wave subject to viscous overstability is predicted to travel indefinitely with an asymptotically constant amplitude, limited 
by nonlinearity.
This behavior is also predicted by the streamline model of \citet{bgt1986}.
In our new model this saturation results from the nonlinear terms in the hydrodynamic balance equations and is not captured by the linearized theory.
Nonlinear terms influence the density wave damping as soon as the perturbation of surface density attains a significant fraction of its background value,
and their magnitude is a measure for the nonlinearity of the wave.

The paper is organized as follows. In Section \ref{sec:lintheo} we briefly review important aspects of the linear theory of density waves, particularly the 
damping by viscosity. 
Section \ref{sec:hydropar} provides detailed information on the numerical values used for the parameters of our hydrodynamical model.
The model equations for the description of a dense ring are presented in Section \ref{sec:hydro}. 
In Section \ref{sec:nlampeq} we use the results of the linear theory in combination with symmetry arguments to place restrictions on the shape of the amplitude 
equation. These considerations motivate the definition of parameters which will then be used for a rigorous 
derivation of the nonlinear amplitude equation in terms of a multiple scale expansion
(Sections \ref{sec:msexp}-\ref{sec:ampeq}). Further, in Sections \ref{sec:ampeq} and \ref{sec:wkbsg} we discuss some general implications of the resulting 
amplitude equation.
In section \ref{sec:nlforced} we include the effects of satellite forcing in the model and derive a 
forced nonlinear amplitude equation. In Section \ref{sec:bgtcomp} we compare our model with the nonlinear streamline model of \citet{bgt1986}. 
Section \ref{sec:summary} provides a summary of the derived results and addresses
remaining open questions.


\section{Linear Viscous Theory}\label{sec:lintheo}

In the linear hydrodynamic theory (\citet{goldreich1978b};~\citet{shu1984}), a trailing density wave 
in the vicinity of an inner Lindblad resonance is described in terms of the perturbed surface mass density 
 \begin{equation}\label{eq:sigmaperturb}
\sigma(r,\theta,t) = \sigma_{0} + \text{Re}[ A\left(r\right) \cdot \exp\Bigg\{i \int_{}^{r}k \left( s\right)\, \mathrm{d}s \Bigg\}  \cdot 
\exp\left\{i\left(m\theta-\omega t\right) \right\}],
\end{equation} 
where $\text{Re}[\,]$ denotes the real part and where $\sigma_{0}$ is the unperturbed surface mass density.
Here we adopt a cylindrical coordinate system ($r$,$\theta$,$z$), restricted to the plane $z=0$.
We consider waves which are excited by single Fourier modes of an external satellite potential, which can be written in the form 
 \begin{equation}\label{eq:phisat}
\phi_{s}\left(r,\theta,t\right)   = \hat{\phi}_{s}\left(r\right) \exp\left\{i\left(m\theta-\omega t\right) \right\},
\end{equation} 
where the radial function $\hat{\phi}_{s}\left(r\right)$ varies slowly with radius $r$ (see \citet{goldreich1980} for more details).
In this paper we neglect higher moments of Saturn's mass distribution, so that the orbital frequencies of ring particles are identical to their corresponding 
epicyclic frequencies and frequencies of vertical oscillations, i.e. $\Omega\left(r\right) =\kappa\left(r\right)=\mu\left(r\right)$
with 
\begin{equation}\label{eq:kepler}
 \Omega\left(r\right) = \sqrt{\frac{G M_{p}}{r^3}}
\end{equation}
where $M_{p}$ denotes Saturn's mass and $G$ is the gravitational constant.
Then the relationship between the Lindblad resonance radius $r_{L}$ and the forcing frequency $\omega$ is
 \begin{equation}\label{eq:rldef}
\omega= (m-1) \,\Omega \left(r_{L} \right).
\end{equation} 
For simplicity we assume that the satellite is on an uninclined, circular orbit so that vertical and corotation resonances can be ignored.
Then, the forcing frequency is simply
 \begin{equation}
\omega= m\, \Omega_{s},
\end{equation} 
where $m$ is the same positive integer as in (\ref{eq:sigmaperturb}) and $\Omega_{s}$ denotes the satellite's orbital frequency.
Since the satellite potential (\ref{eq:phisat}) varies slowly in $r$-direction one typically evaluates the corresponding forcing terms in
the evolutionary equations (Section \ref{sec:nlforced}) at the resonance $r=r_{L}$.
For all times $t$ the perturbation (\ref{eq:sigmaperturb}) is a tightly wound $m$-armed spiral wave which satisfies the WKB-approximation, such that $|k A| \gg 
\partial_{r}A$, meaning that its phase varies much faster with radial distance than its amplitude.
The excitation of the density wave takes place within a small region around the ILR of approximately one wavelength.

The complex wavenumber of the density wave in (\ref{eq:sigmaperturb}) reads (\citet{goldreich1978b}; \citet{shu1984})
 \begin{equation}\label{eq:gt78}
k = \frac{\mathcal{D}x}{2 \pi G \sigma_{0}}+i \,\frac{\Omega_{L} \, \mathcal{D}^2 x^{2}}{(2\pi G \sigma_{0})^3}\, \frac{\eta_{0}}{\sigma_{0}} 
\,\left[\frac{7}{3} + \gamma \right]
\end{equation} 
with
 \begin{equation}\label{eq:xx}
x=\frac{r-r_{L}}{r_{L}}
\end{equation} 
being the scaled distance from the resonance radius $r_{L}$. 
Further, $\mathcal{D}=3\left(m-1\right)\Omega_{L}^2$ (\citet{cuzzi1984}) with $\Omega_{L}=\Omega\left(r_{L}\right)$ being the orbital frequency at resonance. 
In the expression for $k$ the real part yields the linear wavelength dispersion of the wave, while the imaginary part accounts for the effects of viscosity on 
the wave amplitude. 
$\eta_{0}$ is the constant dynamic shear viscosity and the constant $\gamma$ denotes
the ratio of the bulk and shear viscosities.
If one allows for a density dependence of the dynamic viscosity (\citet{schmit1995})
 \begin{equation}\label{eq:shearvis}
\eta = \nu_{0} \, \sigma_{0}  \left(\frac{\sigma}{\sigma_{0}}\right)^{\beta+1},
\end{equation} 
where $\nu_{0}$ denotes the ground state value of the kinematic shear viscosity $\nu=\eta/\sigma$, one obtains (\citet{schmidt2016}) 
possible values of $\sigma$ about the groundstate value one obtains (\citet{schmidt2016})
 \begin{equation}\label{eq:klin}
k  = \frac{\mathcal{D}x}{2 \pi G \sigma_{0}}+i \,\frac{\Omega_{L} \, \mathcal{D}^2 x^{2}}{(2\pi G \sigma_{0})^3}\,\frac{7}{3}\, \nu_{0} 
\,\left[1-\frac{9}{7}\left(\beta+1\right) +\frac{3}{7}\gamma \right].
\end{equation}
Clearly, expression (\ref{eq:gt78}) is recovered in the limit $\beta \to -1$, corresponding to a constant dynamic viscosity (\ref{eq:shearvis}).
From (\ref{eq:klin}) one obtains an exponentially growing wave amplitude with increasing distance from the resonance if
 \begin{equation}\label{eq:marstabnlo}
\beta > \frac{1}{3}\left(\gamma-\frac{2}{3}\right)\equiv \beta_{c}.
\end{equation} 
This is the hydrodynamic criterion for the onset of viscous overstability (\citet{schmit1995}; \citet{schmidt2001b}) in the long wavelength limit. 
Estimates for the bulk viscosity of a dense ring from N-body simulations (\citet{salo2001}) yield $\gamma \sim 2-4$. This suggests $\beta_{c} \approx 1$ in 
dense rings.
For optical depths $\tau \ge 1$, $\beta$ was found analytically (\citet{araki1986}) and by molecular dynamics simulations 
(\citet{wisdom1988};~\citet{salo1991b}) to be larger than one. Moreover, 
in local N-body simulations, including gravitational inter-particle forces, it was shown that in the presence of self-gravity wakes
the kinematic shear viscosity behaves as (\citet{daisaka2001})
 \begin{equation}\label{eq:daisaka}
\nu \simeq C \frac{G^2 \sigma^2}{\Omega^3},
\end{equation} 
which would imply $\beta=2$. In this formula $G$ is the gravitational constant and $C$ denotes a dimensionless constant that depends on particle size 
and bulk density.

\citet{schmidt2003}, which are listed in Table \ref{tab:scalings}.
vertical oscillations
component of self-gravity would do.
(\citet{salo1992b};~\citet{richardson1994};~\citet{daisaka1999};~\citet{ohtsuki2000}), 
model with the
respect to temperature and surface density, respectively.
of \citet{schmidt2003} by the corrective factor 

\section{Hydrodynamic Parameters}\label{sec:hydropar}
For the numerical solution of the hydrodynamic equations (\ref{eq:nleq}), we must specify several quantities, 
like the pressure and the shear viscosity, along with their dependence on the surface mass density of the ring. 
To this end we will use parameters derived from N-body simulations (\cite{salo2001}), as they were used by \citet{schmidt2003} to describe the nonlinear 
evolution of overstable modes in N-body simulations of a dense ring in terms of a hydrodynamic model.
The specific numbers used in our integrations are listed in Table \ref{tab:scalings}.

The hydrodynamic parameters from \citet{salo2001} were determined from simulations without direct particle-particle self-gravity. 
Instead, effects of self-gravity were mimicked by using an artificially 
increased frequency of vertical oscillations of the ring particles, 
a treatment that was introduced by \cite{wisdom1988}. This modification 
increases the collision frequency between particles in a qualitatively similar 
manner as the vertical component of self-gravity would do, which tends to 
promote overstability. This treatment of self-gravity, however, misses the 
effect of self-gravity wakes, which form in the system as a result of 
gravitational instability 
(\citet{salo1992b};~\citet{richardson1994};~\citet{daisaka1999};~\citet{
ohtsuki2000}). Observational evidence for the presence of self-gravity wakes is 
found in large parts of Saturn's rings 
\citep{colwell2006b,French2007,colwell2007}. Nevertheless, the hydrodynamic 
coefficients determined by \citet{salo2001} are well suited as parameters for 
the numerical study performed in this paper, because our model, as well as any 
other theory for spiral density waves presented in the literature so far (see Section \ref{sec:intro} for references), 
does not yet take into account the effect of self-gravity wakes on the evolution of 
the density wave. How the presence of such micro structure, like self-gravity wakes or 
overstable waves, affects the density waves remains a challenge for future 
modeling. 
 
One problem with the hydrodynamic parameters given by \citet{salo2001} 
is that they depend on the velocity dispersion of the ring particle ensemble, in addition to their dependence on the surface mass density. In contrast, our 
hydrodynamic model (Section \ref{sec:hydro}) is isothermal, i.e.\ it assumes a constant velocity dispersion everywhere in the perturbed ring. But it was shown 
\citep{salo2001,schmidt2001b} that thermal modes play a stabilizing role for the development of overstability in a planetary ring (see also 
\citet{Spahn2000a}), such that the stability boundary for overstability is shifted to higher optical depths when compared to an isothermal treatment. It was 
noted by \citet{schmidt2003} that the effects of the thermal modes on the oscillation frequency and on the growth rate of a linear overstable wave can be 
incorporated easily into a purely isothermal model if one renormalizes two hydrodynamic parameters appropriately. These parameters are the derivative of 
pressure with respect to surface mass density ($p_\sigma$) and the ratio of the bulk and shear viscosities ($\gamma$ in this paper). The renormalization uses 
the non-isothermal transport coefficients determined by \citet{salo2001} and the linear non-isothermal mode analysis by \citet{schmidt2001b}. Specifically, it 
is achieved by absorbing in equation (24) of \citet{schmidt2001b} the quantity $F_2$ into an effective $p_\sigma$ and by absorbing the quantity $F_3$ into an 
effective value for the constant ratio $\gamma$. This method led to a quantitative match of an isothermal model for the nonlinear evolution of viscously 
overstable modes with N-body simulations (\cite{schmidt2003}). We will use the same method to fix the parameters of our isothermal model for density waves.

The specific parameter sets used in this paper are listed in Table \ref{tab:scalings}.
These correspond directly to the numbers given in Table (1) of \citet{schmidt2003}, taking into account the different scalings that were applied to 
non-dimensionalize the parameters (see captions of the tables). The parameter $p_\sigma$ in Table \ref{tab:scalings} corresponds to the effective quantity in 
the table by \citet{schmidt2003}. Similarly, our parameter $\gamma$, describing the ratio of bulk to shear viscosity, relates to the parameter 
$\alpha^{\mbox{eff}}$ in \citet{schmidt2003} through $\gamma = \alpha^{\mbox{eff}}-4/3$.

\vspace{ -0.cm}
\begin{minipage}{\textwidth}
\centering
\captionof{table}{\textbf{Values for Parameters and their Scaling}}\label{tab:scalings} 
   \renewcommand*{\arraystretch}{0.9}
    \begin{tabular}{|l|l|l|l|l|l|}
 \hline
    Quantity & Scaling & \multicolumn{4}{|c|}{(Typical) Value} \\     \hline
            $\epsilon=\frac{2\pi G \sigma_{0} }{r_{L} \mathcal{D}}$ (dimensionless parameter)  &  & \multicolumn{4}{|c|}{$10^{-8}-10^{-9}$}\\  
    $G$ (gravitational constant) &  &   \multicolumn{4}{|c|}{$6.67 \cdot 10^{-11} \, \text{m}^3 \, \text{kg}^{-1} \text{s}^{-2}$}\\ 
        $r_{L}$ (resonance radius) & &  \multicolumn{4}{|c|}{$10^8 \, \text{m}$}\\ 
        $\Omega_{L}$ (orbital frequency at resonance) &  & \multicolumn{4}{|c|}{$2 \cdot 10^{-4}\, \text{s}^{-1}$}\\
        $\sigma_{0}$ (ground state surface density) &  & \multicolumn{4}{|c|}{}\\ \hline
    $t$ (time) & $ \Omega_{L}^{-1} $ &   \multicolumn{4}{|c|}{}\\ 
    $k$ (wavenumber) & $ \epsilon^{-1}\,  r_{L}^{-1} $ &  \multicolumn{4}{|c|}{}\\ 
    $u$, $v$ (planar velocity components ) & $ \epsilon\,  r_{L} \, \Omega_{L}$ &  \multicolumn{4}{|c|}{}\\
     $\phi$, $\phi_{s}$ (gravitational potentials) & $\epsilon^2 \, r_{L}^{2} \, \Omega_{L}^{2}$ &   \multicolumn{4}{|c|}{}\\
     		$p$ (scalar pressure) & $\sigma_{0} \, \epsilon^2 \, r_{L}^{2} \, \Omega_{L}^{2}$ &  \multicolumn{4}{|c|}{}\\

		$\sigma$ (surface mass density) & $\sigma_{0}$ &  \multicolumn{4}{|c|}{}\\
		$\Omega$ (orbital frequency) & $\Omega_{L}$ &  \multicolumn{4}{|c|}{}\\ 
		$\omega$ (forcing frequency) & $\Omega_{L}$ &  \multicolumn{4}{|c|}{}\\  
		$\mathcal{D}=3\left(m-1\right)\Omega_{L}^2$ & $\Omega_{L}^2$ &  \multicolumn{4}{|c|}{}\\  \hline \hline
                \multicolumn{6}{|c|}{Hydrodynamic parameters (from \citet{schmidt2003})}\\ \hline
                $\tau$ (optical depth) &  & 1.0 ($\tau_{10}$) & 1.4 ($\tau_{14}$) & 1.5 ($\tau_{15}$) & 2.0 ($\tau_{20}$)  \\ 	

		$\nu_{0} \, [10^{-4}\,  \text{m}^2 \,  \text{s}^{-1}]$  & $\epsilon^2 \,r_{L}^2 \, \Omega_{L} $  & $ 4.43 $ & $ 6.06   $& $ 6.47  $& $ 8.93  $\\
		$\beta_{c}$ & & 1.23 & 0.97  & 0.93 & 0.92 \\
		$\beta$  & & 0.85  &  1.03 & 1.06 & 1.16  \\  
		$\gamma$  & & 4.37  & 3.59 & 3.47 & 3.42  \\    	
		$\delta_{\nu}$ &  & $0.56 \, i$ &  0.25 & 0.37 & 0.51  \\
		$p_{\sigma}  \, [10^{-6} \, \text{m}^{2}\, \text{s}^{-2}]$  & $\epsilon^2 \,  \,r_{L}^2\, \Omega_{L}^2$  & $  0.52  $ & $  
0.63  $ & $ 0.67 $  & $ 1.00 $ \\ \hline	
    \end{tabular}\par
   \bigskip
   \begin{flushleft}
    Note: the quantity $\delta_{\nu}$ is imaginary for $\tau=1.0$ which 
follows from Eq. (\ref{eq:expar}). Further, $\beta_{c}$ is defined in Eq. (\ref{eq:marstabnlo}). For explanations of the hydrodynamic parameters see Sections 
\ref{sec:lintheo}, \ref{sec:hydropar}, \ref{sec:hydro} and \ref{sec:msexp}.
   \end{flushleft}
\end{minipage}

\FloatBarrier
\newpage

\section{Nonlinear Damping of Free Density Waves}\label{sec:nlfree}

In the following we study the influence of nonlinearities in the hydrodynamic equations (\ref{eq:nleq}) on the propagation of density waves in a ring region 
which is described by the viscosity model (\ref{eq:shearvis}) and 
that may exhibit viscous overstability. We formally restrict our considerations to the \textit{weakly nonlinear} regime. Strictly, this means that we are 
sufficiently close to the threshold for the instability (\ref{eq:marstabnlo}) [i.e.  $|\frac{\beta-\beta_{c}}{\beta_{c}}|\ll1 $] and that the density wave has a 
small initial amplitude at the resonance 
location\footnote{such that the density perturbations are much smaller than the equilibrium value: $|\frac{\sigma\left(r\right)-\sigma_{0}}{\sigma_{0}}| \ll 
1$.}. 
It is then appropriate (\citet{cross1993}) to calculate the nonlinear pattern in terms of a multiple scale expansion about the marginally unstable (or 
marginally stable) wave of the linear theory [cf. (\ref{eq:sigmaperturb})]: 
 \begin{equation}\label{eq:scalesep}
\sigma = \sigma_{0} + \text{Re}[\mathcal{A}\left(\xi \right) \cdot \exp\Bigg\{i \int_{}^{x}k\left(s\right)\, \mathrm{d}s \Bigg\}  \cdot 
\exp\left\{i\left(m\theta-\omega t\right) \right\}] +\text{hh}.
\end{equation} 
The amplitude $\mathcal{A}$ will now depend on a ``slow'' radial length scale $\xi$ (formally much larger than one wavelength) and is governed by a Landau-type 
nonlinear amplitude equation.
The wave will accordingly develop nonlinear properties, including the excitation of its higher harmonics (hh). 
The amplitude equation which will be derived below is a nonlinear generalization of the linear damping relation in the case of density dependent viscosities 
(Eq. \ref{eq:klin}).

\subsection{Hydrodynamic Equations}\label{sec:hydro}

We use the cylindrical coordinate system $(x,\theta,z)$ in the plane $z=0$ with the dimensionless distance $x$ as defined in Section \ref{sec:lintheo}. We scale 
length with $r_{L} \epsilon$, where the small dimensionless parameter
 \begin{equation}\label{eq:epsilon}
\epsilon = \frac{2 \pi G \sigma_{0}}{r_{L} \mathcal{D}}
\end{equation}  
describes the strength of self-gravity as compared with the gravity of the central planet.
Expressed in terms of the Toomre critical wavelength $\lambda_{cr}$ this length scale yields typical values $r_{L} \epsilon \sim \lambda_{cr} / 6 \pi \sim 2 \, 
\text{m}$. 
Further, time is scaled with $1/\Omega_{L}$ and surface density with its ground state value $\sigma_{0}$. 
The scaled $z$-integrated nonlinear isothermal fluid equations in the plane $(z=0)$ then read (\citet{stewart1984};~\citet{schmidt2009})
 \begin{align}\label{eq:nleq}
\begin{split}
\partial_{t} \sigma & = - \Omega(x) \, \partial_{\theta}\sigma- \epsilon \left(\sigma \, \partial_{x} u + u\, \partial_{x} \sigma \right), \\[0.1cm]
 \partial_{t} u  &  = -\Omega(x) \,\partial_{\theta}u  + 2 \Omega(x) \, v -\epsilon \, u \, \partial_{x} u\\
 &  \quad   + \nu_{0} \, \epsilon^2 \left(\frac{4}{3}+\gamma \right) \left(1+\beta \right) \sigma^{\beta-1} \partial_{x} \sigma \partial_{x} u \\ 
&  \quad  +\nu_{0} \,\epsilon^2 \left(\frac{4}{3}+\gamma \right) \, \sigma^{\beta}\partial_{x}^{2} u- \epsilon \, p_{\sigma}  \frac{\partial_{x} 
\sigma}{\sigma}-\epsilon \, \partial_{x} \phi -\epsilon \, \partial_{x} \phi_{s}, \\[0.1cm]
\partial_{t} v & = -\Omega(x) \, \partial_{\theta}v -\frac{1}{2}\Omega\left(x\right) u- \epsilon \, u\,\partial_{x} v\\
&  \quad  +  \nu_{0}\, \epsilon \left(1+\beta\right) \sigma^{\beta -1} \partial_{x} \sigma \left(\epsilon \, \partial_{x} v-\frac{3\Omega(x)}{2}\right)\\
&  \quad  + \nu_{0} \,\epsilon^2 \, \sigma^{\beta} \, \partial_{x}^{2} v -\epsilon \, \partial_{\theta}\phi_{s}.
\end{split}
\end{align} 
The symbols $\sigma$, $u$ and $v$ denote the surface mass density, the radial and the tangential velocities, respectively. Note that $v$ does not include the 
Keplerian ground state velocity $\Omega  r$ [cf. (\ref{eq:kepler})]
and that we neglect curvature terms, since we will focus on the description of tightly wrapped waves whose wavelengths fulfill $\lambda \ll r$.
We use the viscosity prescription (\ref{eq:shearvis}). 
In the equation for the radial velocity $u$ the derivative of the scalar pressure $p$ is approximated as
\begin{equation}
 \frac{\mathrm{d}p}{\mathrm{d}x} = p_{\sigma} \frac{\partial \sigma}{\partial x}
\end{equation}
with
\begin{equation}
 p_{\sigma} \equiv \left[\frac{\partial p}{\partial \sigma}\right]_{0}
\end{equation}
where the subscript ``0'' denotes that the derivative has to be taken at the ground state. The quantity $p_{\sigma}$ is scaled with $\left(\epsilon \, r_{L} 
\Omega_{L} \right)^2$.
Using values for $p_{\sigma}$ from simulations (listed in Table \ref{tab:scalings}) this linearized treatment of the equation of state retains
effects of non-local pressure.
Further, the quantities $\phi$ and $\phi_{s}$ are the self-gravity potential and the satellite potential, respectively.
From here on, all parameters and quantities are scaled as denoted in Table \ref{tab:scalings}.

Equations (\ref{eq:nleq}) are vertically averaged. This restricts their applicability to phenomena which occur on radial length scales much greater than the 
vertical extent of the disk.
For density waves this condition is fulfilled by a large margin. However, it is expected that in regions of high compression, such as the peaks of density 
waves, vertical splashing of the ring material occurs, similar as in overstable oscillations (\citet{salo2001}). 
In these regions the isothermal approximation is violated. Qualitatively, one would expect an increased velocity dispersion in regions of higher compression,
such that, to first order, the increased energy in the random motions gets balanced by an enhanced frequency of inelastic particle collisions. 
Keplerian.
The effect of the satellite forcing terms in Eqs. (\ref{eq:nleq}) will be studied in Section \ref{sec:nlforced}. However, in the free wave analysis which 
follows below we exclude these terms.
We restrict our analysis to \textit{long trailing density waves near an inner Lindblad resonance} such that $x \ll 1$.

\subsection{Nonlinear Amplitude Equation}\label{sec:nlampeq}

Our aim is to derive a \textit{complex nonlinear amplitude equation} for the radial steady state profile of a density wave, propagating away from an ILR. 
Before we proceed with a rigorous derivation in Section \ref{sec:msexp}, we can already place certain restrictions on the shape of this equation by means of 
physical arguments and by using the results of linear theory (Section \ref{sec:lintheo}).

First of all, since the wave amplitude $\mathcal{A}$ is time independent in a stationary state, time derivatives shall not appear.
Further, the equation should be invariant upon multiplying $\mathcal{A}$ by an arbitrary phase factor. 
This can be seen by applying the multiplication
$\mathcal{A} \rightarrow \mathcal{A}\cdot \exp\left(i\Phi\right)$ to (\ref{eq:scalesep}) which describes the density wave state in the lowest approximation of 
the order parameter expansion which follows below. One sees that it is always possible to absorb the phase factor $\exp\left(i\Phi\right)$ in the phase $ i 
\omega t$,
which corresponds to a translation in time and the amplitude equation must be invariant upon this translation. Therefore, the simplest possible nonlinear 
amplitude equation has the form
 \begin{equation}\label{eq:landaumodel}
\frac{\mathrm{d} \mathcal{A}}{\mathrm{d}x} = g\left(x\right) \mathcal{A} - l\left(x\right) \mathcal{A}|\mathcal{A}|^{2},
\end{equation}
where nonlinear effects are described through the cubic term and where the radial dependence of the amplitude is written in terms of the regular length scale 
$x$.
The applied sign convention resembles the one which is commonly used when formulating the \emph{complex Ginzburg-Landau equation} (\citet{aranson2002}).
In general the functions $g\left(x\right)$ and $l\left(x\right)$ are complex and their dependence on $x$ reflects the fact that the considered system is 
\emph{not} invariant upon translation
in $x$-direction, since, to lowest order in $x$, the wavenumber of the density wave depends linearly on $x$ [cf. (\ref{eq:klin})]. 

In the \textit{linear limit}, obtained for sufficiently small values of $\mathcal{A}$, the nonlinear term in (\ref{eq:landaumodel}) is negligible and the 
equation should reproduce the linear damping relation described by the \emph{imaginary part} $k_{i}$ of the complex wavenumber (\ref{eq:klin}).
In this limit, integration of (\ref{eq:landaumodel}) yields
 \begin{equation}\label{eq:alin1}
\mathcal{A}(x)=\exp\left(\int_{}^{x} g\left(s\right) \mathrm{d}s \right).
\end{equation} 
On the other hand, if we insert the imaginary part $k_{i}$ of the wavenumber (\ref{eq:klin}) in (\ref{eq:sigmaperturb}) and apply the scaling discussed in 
Section \ref{sec:hydro}, we obtain
 \begin{equation}\label{eq:alin2}
\mathcal{A}(x)=\exp\left(\int_{}^{x}      \frac{3\,\nu_{0} \,s^{2} }{ \epsilon \mathcal{D}}\left(\beta -\beta_{c}\right)  \, \mathrm{d}s\right),
\end{equation} 
where the quantities $\nu_{0}$ and $\mathcal{D}$ are scaled.
Comparison of (\ref{eq:alin1}) with (\ref{eq:alin2}) leads to the condition
 \begin{equation}
\text{Re}[g\left(x\right)]= \frac{3\,\nu_{0} \,x^{2} }{ \epsilon \mathcal{D} } \left(\beta -\beta_{c}\right).
\end{equation} 
This can be rewritten in the form
 \begin{equation}\label{eq:grsoll}
g_{r}\left(x\right)= \frac{\,\nu_{0} \,x^{2} \left( 3\gamma -2\right) }{ 3\epsilon \mathcal{D} } \left( \frac{\beta -\beta_{c}}{\beta_{c}}\right)
\end{equation} 
where we defined $g_{r}\left(x\right) =\text{Re}[g\left(x\right)]$. The last expression clearly displays the role of the viscous parameter $\beta$ as a 
threshold parameter for the linear instability of a density wave as it occurs for $\beta>\beta_{c}$.
In the same manner $\beta$ is used as threshold parameter in the linear theory of viscous overstability (\citet{schmit1995}).

In the case $\beta>\beta_{c}$ the wave amplitude (\ref{eq:alin2}) grows exponentially. This means that the linear description fails and the cubic term in 
(\ref{eq:landaumodel}) is \textit{necessary} to provide a damping of the wave. 
In the following sections we perform a multi-scale expansion of the hydrodynamic equations (\ref{eq:nleq}) in order to \emph{derive} Eq. 
(\ref{eq:landaumodel}), where the linear coefficient $g(x)$ will be identical to (\ref{eq:grsoll}), consistent with the linear theory.

\subsection{Multiple Scale Expansion}\label{sec:msexp}

In order to perform the multiple scale expansion, we define as control parameter for the bifurcation from the ground state 
 \begin{equation}\label{eq:expar}
\delta_{\nu} = \sqrt{\frac{\beta - \beta_{c}}{\beta_{c}}},
\end{equation} 
such that the bifurcation occurs at $\delta_{\nu}=0$. 
The subscript $\nu$ is used to distinguish this expansion parameter from the small parameter $\delta_{s}$ which will be introduced in Section 
\ref{sec:nlforced} to describe forced density waves.
From this definition directly follows that the linear coefficient (\ref{eq:grsoll}), and consequently the imaginary part of the wavenumber (\ref{eq:klin}), are 
both proportional to $\delta_{\nu}^{2}$
such that these quantities change their signs at the bifurcation which marks the threshold for linear instability.

The state variables are expanded as a series in powers of $|\delta_{\nu}|$:
\begin{subequations}\label{sub:msexp}
 \begin{align}
\phi & = |\delta_{\nu}| \, \phi_{1} + |\delta_{\nu}|^{2} \, \phi_{2} + |\delta_{\nu}|^{3}\, \phi_{3} + \cdots \hspace{0.1 cm}, \label{sub:phiexp}\\
u & = |\delta_{\nu}| \, u_{1} + |\delta_{\nu}|^{2} \, u_{2} + |\delta_{\nu}|^{3} \, u_{3} + \cdots \hspace{0.1 cm}, \label{sub:uexp}\\
v & = |\delta_{\nu}| \, v_{1} + |\delta_{\nu}|^{2} \, v_{2} + |\delta_{\nu}|^{3} \,  v_{3} + \cdots \hspace{0.1 cm}, \label{sub:vexp}\\
\sigma & = 1+ |\delta_{\nu}| \, \sigma_{1} + |\delta_{\nu}|^{2} \, \sigma_{2} + |\delta_{\nu}|^{3}\,  \sigma_{3} + \cdots \hspace{0.1 cm}, 
\label{sub:sigmaexp}\\
\beta & = \beta_{c}+|\delta_{\nu}| \, \beta_{1} + |\delta_{\nu}|^{2} \, \beta_{2} + \cdots\label{sub:betaexp}\hspace{0.1 cm}.
\end{align} 
\end{subequations}
Further, we introduce a ``slow radial length scale'' $\xi$ by
 \begin{equation}\label{eq:slowlength}
\partial_{x}  \rightarrow \partial_{x} + |\delta_{\nu}|^{2} \partial_{\xi}.
\end{equation} 
We use the absolute value $|\delta_{\nu}|$ for the expansion. This is done as to avoid a negative slow length scale which would otherwise occur if 
$\beta<\beta_{c}$ since for this case $\delta_{\nu}^2 <0$.
By using the absolute value we further ensure that all terms in the expressions (\ref{sub:phiexp})-(\ref{sub:betaexp}) are real-valued.
The necessary expansion of $\beta$ arises from the choice of the expansion parameter (\ref{eq:expar}). The consistency of the expansion requires that finally
 \begin{equation}\label{eq:betaexp}
\beta = \beta_{c}\left(1+\delta_{\nu}^{2}\right),
\end{equation}
such that $\beta$ is a real quantity and can take values greater or smaller than $\beta_{c}$.
With (\ref{sub:betaexp}) this implies the conditions
\begin{subequations}
\begin{align} 
\beta_{1} & = 0\hspace{0.1 cm},\label{sub:beta1}\\
\beta_{2} & = \beta_{c} \, \mathrm{sgn}\left(\delta_{\nu}^2\right) \hspace{0.1 cm}, \label{sub:beta2}
\end{align}
\end{subequations}
where $\mathrm{sgn}\left(\delta_{\nu}^2\right)$ denotes the sign of $\delta_{\nu}^2$.

The two length scales $x$ and $\xi$ are well separated for small values of $|\delta_{\nu}|$, i.e. near the threshold for instability. The ``fast'' scale $x$ 
describes the radial oscillation
of the density wave, defined by the real wavenumber, which will be written in the following as $k\left(x\right)$. In contrast, the variation of the wave 
amplitude occurs on the ``slow'' scale $\xi$.
For the vector of state in the plane ($z=0$) we adopt the short notation
 \begin{equation}\label{eq:msexpvec}
\mathbf{\Psi}  = \sum\limits_{i}^{} |\delta_{\nu}|^{i} \, \mathbf{\Psi}_{i}\left(x,\theta,t, \xi \right)
\end{equation} 
with
 \begin{equation}\label{eq:msexpvec2}
\mathbf{\Psi}_{i}\left(x,\theta,t,\xi\right) = \begin{pmatrix} \phi_{i}\left(x,\theta,t,\xi\right)\\ u_{i}\left(x,\theta,t,\xi\right)\\ 
v_{i}\left(x,\theta,t,\xi\right) \end{pmatrix},
\end{equation} 
where the self-gravity potential is used in place of the surface mass density $\sigma$ to describe the hydrodynamic state. This means that the surface density 
in Eqs. (\ref{eq:nleq}) has to be replaced by a solution of Poisson's equation which is derived in Appendix \ref{sec:poisson}.
Inserting (\ref{eq:slowlength}), (\ref{eq:betaexp}) and (\ref{eq:msexpvec}) in the nonlinear evolution equations (\ref{eq:nleq}), and sorting by orders in 
$|\delta_{\nu}|$ one obtains the following hierarchy of balances when requiring that
all orders of $|\delta_{\nu}|$ vanish separately:
 \begin{equation}\label{eq:eqnexp}
\begin{split}
\mathcal{O}\left(|\delta_{\nu}|^{1}\right): \hspace{0.5 cm} \mathit{\hat{L}} \mathbf{\Psi_{1}} & = 0, \\
\mathcal{O}\left(|\delta_{\nu}|^{2}\right): \hspace{0.5 cm} \mathit{\hat{L}} \mathbf{\Psi_{2}} & = \mathbf{N_{2}}(\mathbf{\Psi_{1}},\mathbf{\Psi_{1}}),\\
\mathcal{O}\left(|\delta_{\nu}|^{3}\right): \hspace{0.5 cm} \mathit{\hat{L}} \mathbf{\Psi_{3}} & = \mathbf{N_{3}}(\mathbf{\Psi_{1}},\mathbf{\Psi_{2}}) + 
\partial_{\xi}(\mathit{\hat{M}}\cdot \mathbf{\Psi_{1}}).
\end{split}
\end{equation}
In the following it is shown that the desired amplitude equation (\ref{eq:landaumodel}) can be obtained from the $\mathcal{O}\left(|\delta_{\nu}|^{3}\right)$ 
equations.
Therefore we truncate the expansion at this order.
In the above equations the same linear operator $\mathit{\hat{L}}$ appears on the left hand side at each order of $\delta_{\nu}$. $\mathit{\hat{L}}$ itself is 
of order $|\delta_{\nu}|^{0}$.
The vectorial terms $\mathbf{N_{i}}$ are the nonlinear terms at order $|\delta_{\nu}|^{i}$ and are provided in Appendix \ref{sec:nlterms}.
With this expansion we have split the problem in an iterative hierarchy of linear inhomogeneous equations where
the inhomogeneities at a given order are functions of the solutions of the lower order equations.
$\mathit{\hat{M}}$ denotes a $3\mathbf{\times} 3$-matrix containing complex constants of order $|\delta_{\nu}|^{0}$.
The operator $\mathit{\hat{L}}$ and its adjoint $\mathit{\hat{L}}^{\dagger}$ are given by (\ref{eq:linop}) and (\ref{eq:linopad}), respectively.
Note that the terms $\mathrm{d}\sigma/\mathrm{d}\phi$ will in general be different for each order in $|\delta_{\nu}|$ (cf. Appendix \ref{sec:poisson}).
 \begin{equation}\label{eq:linop}
\begin{array}{@{}*{22}{l@{}}}
\mathit{\hat{L}} = \begin{pmatrix}\frac{\mathrm{d}\sigma}{\mathrm{d}\phi}\left(\partial_{t}+ \Omega \,\partial_{\theta}\right) \hspace{0.4 cm} & \epsilon \, 
\partial_{x} \hspace{0.4 cm} & 0 \\[0.5 cm] 
\epsilon \left(1+p_{\sigma} \frac{\mathrm{d}\sigma}{\mathrm{d} \phi}\right)\partial_{x}  \hspace{0.4 cm} & \partial_{t}+\Omega 
\,\partial_{\theta}-\left(\frac{4}{3}+\gamma \right)\nu_{0}\, \epsilon^2 \,  \partial_{x}^{2} \hspace{0.4 cm}  & -2 \Omega \\[0.5 cm] 
\frac{3}{2}\Omega\,  \nu_{0}\left(1+\beta_{c}\right)\frac{\mathrm{d}\sigma}{\mathrm{d}\phi}\epsilon \,\partial_{x} \hspace{0.4 cm} &\dfrac{1}{2}\Omega 
\hspace{0.4 cm}&  \partial_{t}+\Omega \, \partial_{\theta}-\nu_{0}\, \epsilon^2 \, \partial_{x}^{2}\end{pmatrix} .
\end{array}
\end{equation} 
\vspace{0.1cm}
 \begin{equation}\label{eq:linopad}
\begin{array}{@{}*{22}{l@{}}}
\mathit{\hat{L}}^{\dagger} = \begin{pmatrix}   -\frac{\mathrm{d}\sigma}{\mathrm{d}\phi}\left(\partial_{t}+ \Omega \,\partial_{\theta}\right) \hspace{0.4 cm} &  
-\left(1+p_{\sigma} \frac{\mathrm{d}\sigma}{\mathrm{d \phi}}\right)\epsilon \, \partial_{x} \hspace{0.4 cm} & -\frac{3}{2}\Omega\,  
\nu_{0}\left(1+\beta_{c}\right)\frac{\mathrm{d}\sigma}{\mathrm{d}\phi}\epsilon \,\partial_{x} \\[0.5 cm]- \epsilon \, \partial_{x}  \hspace{0.4 cm} & 
-\partial_{t} - \Omega \,\partial_{\theta}-\left(\frac{4}{3}+\gamma \right)  \nu_{0}\, \epsilon^2 \, \partial_{x}^{2} \hspace{0.4 cm}  & \dfrac{1}{2}\Omega 
\\[0.5 cm] 0 \hspace{0.4 cm} & -2 \Omega \hspace{0.4 cm}&  -\partial_{t}-\Omega \, \partial_{\theta}-\nu_{0}\, \epsilon^2 \, \partial_{x}^{2}\end{pmatrix} .
\end{array}
\end{equation} 
To solve (\ref{eq:eqnexp}) we have to apply appropriate solvability conditions\footnote{This is necessary to avoid resonant driving of the linear equations 
(\citet{cross1993}).}.
Therefore a scalar product has to be defined in the space of the vectors (\ref{eq:msexpvec2}) [whose components are of the form (\ref{eq:scalesep})]. 
If $\mathit{\hat{L}}^{\dagger}$ is the adjoint operator of $\mathit{\hat{L}}$ and $\mathbf{\Psi_{0}}^{ad}$ one of its null solutions, i.e.
 \begin{equation}\label{eq:adjointeq} 
\mathit{\hat{L}}^{\dagger} \mathbf{\Psi_{0}}^{ad} =0,
\end{equation} 
then taking the scalar product of the expanded Eqs. (\ref{eq:eqnexp}) with $\mathbf{\Psi_{0}}^{ad}$ leads to the solvability conditions
 \begin{equation}\label{eq:solvcon}
\begin{split}
\mathcal{O}\left(|\delta_{\nu}|^{2}\right): \hspace{0.5 cm} & \langle \mathbf{\Psi_{0}}^{ad}|\mathbf{N_{2}}(\mathbf{\Psi_{1}},\mathbf{\Psi_{1}}) \rangle  = 0,\\
\mathcal{O}\left(|\delta_{\nu}|^{3}\right): \hspace{0.5 cm} & \langle \mathbf{\Psi_{0}}^{ad}|\mathbf{N_{3}}(\mathbf{\Psi_{1}},\mathbf{\Psi_{2}})  + 
\partial_{\xi}(\mathit{\hat{M}}\mathbf{\Psi_{1}}) \rangle  = 0.
\end{split}
\end{equation} 
The solvability condition at $\mathcal{O}\left(|\delta_{\nu}|^{3}\right)$ yields the desired differential equation (\ref{eq:landaumodel}) for the complex 
amplitude $\mathcal{A}$
which is the final goal of the weakly nonlinear analysis. 
As a scalar product we can use
 \begin{equation}\label{eq:scalarprod}
\langle \mathbf{\Psi_{k}}|\mathbf{\Psi_{l}}\rangle \equiv  \frac{1}{2 \pi}\int_{0}^{2 \pi}\mathrm{d} \theta \left[\phi_{k}^{*} \phi_{l}+ u_{k}^{*} u_{l} + 
v_{k}^{*} v_{l}\right],
\end{equation} 
where a star denotes complex conjugate. This choice is appropriate since all fields (\ref{eq:msexpvec}) can be decomposed in exponential phase factors 
$\exp\left\{j  \left[   \int i \frac{k}{\epsilon} \mathrm{d}x + im\theta-i\omega t  \right] \right\}$ with $j=\pm 1,\pm2,\cdots $ such that expressions with 
different exponential phase factors are orthogonal in terms of (\ref{eq:scalarprod}).

\subsection{Linear Stability Problem}\label{sec:linstab}

In this section we solve the order $|\delta_{\nu}|^{1}$ equations (\ref{eq:eqnexp}).
For the order $|\delta_{\nu}|^{1}$ vector of state we assume [cf. (\ref{eq:scalesep})]
 \begin{equation}\label{eq:psi1}
\begin{split}
\mathbf{\Psi_{1}} & = \mathcal{A}(\xi) \, \mathbf{A_{\Psi{1}}}(x) \, \exp\left\{   \int i\, \frac{k}{\epsilon} \mathrm{d}x + im\theta-i\omega t  \right\}\\ 
\quad & +c.c.
\end{split}
\end{equation} 
with the slowly varying amplitude $\mathcal{A}(\xi)$  and where $c.c.$ denotes complex conjugate. Note that since $\mathit{\hat{L}}$ does not act on the slow 
length scale $\xi$,
the amplitude $\mathcal{A}(\xi)$ will be carried along as pre-factor until we arrive at the order $|\delta_{\nu}|^{3}$ equations in Section \ref{sec:ampeq}.
The system of partial differential equations $\mathit{\hat{L}} \mathbf{\Psi_{1}} = 0$ with the differential operator $\mathit{\hat{L}}$, together with 
the ansatz (\ref{eq:psi1}) yield the set of algebraic equations $\hat{L}_{1}\mathbf{A_{\Psi{1}}}(x)=0$ with the complex matrix
 \begin{equation}\label{eq:linop1}
\begin{array}{@{}*{22}{l@{}}}
\hat{L}_{1} \equiv \begin{pmatrix}   -i \left(\omega-m\, \Omega\right) \hspace{0.4 cm} & -i \mathcal{D} \hspace{0.4 cm} & 0 \\[0.5 cm] 
ik\left(1-\frac{p_{\sigma} \, k}{\mathcal{D}}\right) \hspace{0.4 cm} & -i\left(\omega - m \, \Omega\right)+k^{2}\left(\frac{4}{3}+\gamma\right)\nu_{0} 
\hspace{0.4 cm}  & -2 \Omega \\[0.5 cm] -\dfrac{3\Omega \, ik^{2}\left(1+\beta_{c}\right)\nu_{0}  }{2\, \mathcal{D}} \hspace{0.4 cm} &\dfrac{1}{2 }\Omega  
\hspace{0.4 cm}& -i\left(\omega - m \, \Omega\right)+k^{2}\, \nu_{0}\end{pmatrix} 
\end{array}
\end{equation}
and its null space $\mathbf{A_{\Psi{1}}}(x)$.
Below we will see that the $x$-dependency of $\mathbf{A_{\Psi{1}}}(x)$ vanishes in the leading order approximation.
For a null space $\mathbf{A_{\Psi{1}}}(x)$ to exist, the determinant of $\hat{L}_{1}$ must vanish, i.e. 
 \begin{equation}\label{eq:detl1}
\begin{split}
 \mathrm{Det}\,\hat{L}_{1} & =  \left(\omega - m \Omega \right)\left(\frac{D}{\mathcal{D}}-k\right)\\[0.1cm] 
 \quad & + \frac{i\left(-\left(\omega - m \Omega \right)^2\left(7 + 3 \gamma\right) + 9\left(\beta_{c}+1\right)\Omega^2\right)\nu_{0} + 3 
p_{\sigma}\left(\omega- m \Omega\right)}{3 \mathcal{D}}k^2 \\[0.1cm]
 \quad & - i \nu_{0} \,k^3  +\frac{\nu_{0} \left(3ip_{\sigma} + \left(4 + 3 \gamma\right)\left(\omega - m \Omega\right)\nu_{0}\right)}{3 \mathcal{D}}k^4 =0.
\end{split}
\end{equation} 
Here we used the definition
 \begin{equation}\label{eq:dx3}
\Omega^2 - \left(\omega - m \Omega \right)^2 = D 
\end{equation} 
with $D(r_{L})=0$ [cf. (\ref{eq:rldef})] and $r_{L}\left[\mathrm{d}D/\mathrm{d}r\right]_{r_{L}}=\mathcal{D}$.
The condition of marginal stability affords that the wavenumber is real, i.e. $\operatorname{Im}\left[k\left(x\right)\right]=0$.
Therefore, we can solve Eq. (\ref{eq:detl1}) separately for its real and imaginary parts.
The imaginary part yields the exact curve of marginal stability:
 \begin{equation}\label{eq:marstab}
\beta_{c}\left(k\right) = -1+\frac{\left(7 + 3 \gamma\right)\left(\omega - m \Omega\right)^2}{9 \Omega^2} + \frac{\mathcal{D}}{3 \Omega^2}k - 
\frac{p_{\sigma}}{3 \Omega^2}k^2.
\end{equation} 
The condition that the real part of the determinant is zero yields the dispersion relation:
 \begin{equation}\label{eq:xk}
D = \mathcal{D} k - p_{\sigma} \, k^2 - \frac{\left(4+3 \gamma\right)\nu_{0}^2}{3 }k^4.
\end{equation}  
\noindent
If we assume small distances from the Lindblad resonance $x \ll 1$:
 \begin{equation}\label{eq:dexp}
D=\mathcal{D}\, x + \mathcal{O}\left(x^{2}\right).
\end{equation} 
Thus, solving equation (\ref{eq:xk}) perturbatively with the series expansion $k\left(x\right) \equiv k_{0} + k_{1} \, x + k_{2} \, x^2 + \cdots$ yields
the (scaled) wavenumber of marginally stable long trailing
density waves [cf. (\ref{eq:klin})]
 \begin{equation}\label{eq:marginalk}
k = x + \mathcal{O} \left( x^2 \right).
\end{equation} 
We also need to consider the leading order expressions of the frequencies:
\begin{subequations}
 \begin{align}
\Omega & = 1 + \mathcal{O}\left(x\right) \label{sub:freq1},\\
\omega & = \left(m - 1 \right) + \mathcal{O}\left(x\right) \label{sub:freq2}.
\end{align} 
\end{subequations} 
With the above approximations, the curve of marginal stability reduces to relation
 \begin{equation}\label{eq:marstabnlo2}
\beta_{c} = \frac{1}{3}\left(\gamma-\frac{2}{3}\right)+\mathcal{O}\left(x\right).
\end{equation} 
The condition $\beta > \beta_{c}$ is the condition for viscous overstability in the hydrodynamic approximation [cf. (\ref{eq:marstabnlo})]. 

The final goal of the nonlinear analysis is the nonlinear amplitude equation (\ref{eq:landaumodel}) with the proper coefficient functions $g\left(x\right)$ and 
$l\left(x\right)$.
If we carry out our calculations with the relations (\ref{eq:marstab}) and (\ref{eq:xk}), such that the null spaces of $\hat{\mathbf{L}}$ and 
$\hat{\mathbf{L}}^{\dagger}$ exist and can be calculated exactly, 
we can apply the approximations (\ref{eq:marginalk}), (\ref{sub:freq1}, b) at the end and expand corresponding terms to leading order in $x$.
Then, the linear coefficient $g(x)$ in (\ref{eq:landaumodel}) should reduce to the negative of the imaginary part of (\ref{eq:klin}) from linear theory.

The expressions arising in course of the analysis will contain a vast number of terms. We will present here only terms to leading order in $x$ for the sake of 
brevity and clarity.
For the evolution of the amplitude of the waves these terms provide an excellent approximation.

Using (\ref{eq:marstab}) and (\ref{eq:xk}), as well as the approximations (\ref{eq:marginalk}) and (\ref{sub:freq1}, b), the marginal null vector of 
$\hat{L}_{1}$ in the vicinity of an ILR ($x \ll 1$) is given by
 \begin{equation}\label{eq:nullspace}
\begin{array}{@{}*{22}{l@{}}}
\mathbf{A_{\Psi{1}}}(x) = \begin{pmatrix} -2 i  \mathcal{D}  \\[0.1cm] -2  i \\[0.1cm] 1 \end{pmatrix}, 
\end{array}
\end{equation} 
showing no $x$-dependency.
In order to proceed we also need to compute the adjoint null space, i.e. the null space of (\ref{eq:linopad}).
With the ansatz 
 \begin{equation}\label{eq:nullsolad}
\mathbf{\Psi_{0}}^{ad} = \mathbf{A_{\Psi{0}}}^{ad}(x) \, \exp\left\{   \int i\, \frac{k}{\epsilon} \mathrm{d}x + im\theta-i\omega t  \right\}
\end{equation} 
Equation (\ref{eq:adjointeq}) reads $\hat{L}_{1}^{\dagger} \,\mathbf{A_{\Psi{0}}}^{ad}(x) =0$
with the matrix
 \begin{equation}\label{eq:linop1ad}
\begin{array}{@{}*{22}{l@{}}}
\hat{L}_{1}^{\dagger} \equiv \begin{pmatrix}   i\left(\omega - m \Omega\right) \hspace{0.4 cm} & -ik\left(1-\frac{p_{\sigma} k }{\mathcal{D}}\right) \hspace{0.4 
cm} & \dfrac{3 \Omega i k^2 \left(1 + \beta_{c}\right)\nu_{0}}{2 \mathcal{D}} \\[0.5 cm] i\mathcal{D} \hspace{0.4 cm} & i\left(\omega - m 
\Omega\right)+k^{2}\left(\frac{4}{3}+\gamma\right)\nu_{0} \hspace{0.4 cm}  & \dfrac{1}{2}\Omega \\[0.5 cm] 0 \hspace{0.4 cm} & -2 \Omega  \hspace{0.4 cm}& 
i\left(\omega - m \Omega \right) +k^{2}\nu_{0}\end{pmatrix} 
\end{array}
\end{equation} 
\noindent
being the adjoint of (\ref{eq:linop1}). With relations (\ref{eq:marstab}) and (\ref{eq:xk}), the determinant of $\hat{L}_{1}^{\dagger}$ vanishes as well and 
its 
expanded null space reads
 \begin{equation}\label{eq:nullspacead}
\begin{array}{@{}*{22}{l@{}}}
\mathbf{A_{\Psi{0}}}^{ad}(x) = \begin{pmatrix}-\frac{1}{2}\mathcal{D}x \\[0.1cm]-\frac{i}{2} \\[0.1cm] 1 \end{pmatrix}.
\end{array}
\end{equation}

\subsection{Second Order Solution}\label{sec:secondorder}
With the solution for $\mathbf{\Psi_{1}}$ given by (\ref{eq:psi1}), we can proceed to the computation of the second order inhomogeneity 
$\mathbf{N_{2}}(\mathbf{\Psi_{1}},\mathbf{\Psi_{1}})$ (cf. Appendix \ref{sec:nlterms}), the second order solvability condition and the second order vector of 
state $\mathbf{\Psi_{2}}$.
The second order inhomogeneity, expanded to leading order in $x$, reads
 \begin{equation}\label{eq:inhom2}
\begin{split}
& \mathbf{N_{2}}(\mathbf{\Psi_{1}},\mathbf{\Psi_{1}})   =  \begin{pmatrix} - 8i x^2 \\[0.1cm]
4ix \\[0.1cm] 
-2x \end{pmatrix}  \cdot  \exp \left\{2 i\,\left[ \int  \frac{k}{\epsilon} \, \mathrm{d}x-\omega t +m\theta\right] \right\} \mathcal{A}(\xi)^{2} \\[0.1cm]
    \quad & + \begin{pmatrix} 0 \\[0.1cm] 0 \\[0.1cm] 3 \beta_{1} \nu_{0}\, x^2 \end{pmatrix} 
    \cdot \exp \left\{\int  i\, \frac{k}{\epsilon} \, \mathrm{d}x-i\omega t +im\theta\right\} \mathcal{A}(\xi)  \\[0.1cm]
  \quad &  + \begin{pmatrix} 0 \\[0.1cm] -8 \left( \frac{4}{3} + \gamma \right) \nu_{0}\, x^3  \\[0.1cm] 4x \end{pmatrix} \cdot  |\mathcal{A}(\xi)|^{2} \, +c.c. 
\, \, .
\end{split}
\end{equation} 
To obtain a solution for $\mathbf{\Psi}_{2}$ from the $\mathcal{O}\left(|\delta_{\nu}|^{2} \right)$ equation in (\ref{eq:eqnexp}), the corresponding solvability 
condition (\ref{eq:solvcon}) must be fulfilled.
Considering the definition of the scalar product (\ref{eq:scalarprod}) and the adjoint null solution (\ref{eq:nullsolad}), it is clear that after evaluating the 
scalar product, the only remaining terms are those proportional to $\mathcal{A} \left(\xi\right)$ in (\ref{eq:inhom2}).
Thus, the solvability condition reads
 \begin{equation}\label{eq:solvcon2}
3 \beta_{1} \nu_{0} \, x^2 = 0,
\end{equation} 
from which follows:
 \begin{equation}\label{eq:beta1}
 \beta_{1}=0.
\end{equation} 
This is consistent with (\ref{sub:beta1}).
The same relation follows from the exact second order solvability condition, which is also proportional to $\beta_{1}$ and which will not be displayed here, as 
explained above.
With (\ref{eq:beta1}) being satisfied, we are now able to obtain a particular solution for $\mathbf{\Psi}_{2}$.\\
Considering (\ref{eq:inhom2}) and (\ref{eq:beta1}) the particular solution will take the form
 \begin{equation}\label{eq:psi2}
\begin{split}
\mathbf{\Psi_{2,p}} & =\\
\quad & \mathcal{A}(\xi)^{2} \, \mathbf{A_{\Psi{2a}}}(x) \, \exp \left\{2 i\,\left[ \int  \frac{k}{\epsilon} \, \mathrm{d}x-\omega t +m\theta\right] \right\}\\
\quad & + |\mathcal{A}(\xi)|^{2} \, \mathbf{A_{\Psi{2b}}}(x)\\
\quad & +c.c. \hspace{0.5 cm} .
\end{split}
\end{equation} 
Expression (\ref{eq:psi2}) contains six unknown quantities which can be computed by evaluating the three $\mathcal{O}\left(|\delta_{\nu}|^{2}\right)$ equations 
(\ref{eq:eqnexp}) for the oscillatory terms
($\sim  \mathcal{A}(\xi)^{2} \exp \left\{2 i\,\left[ \int  \frac{k}{\epsilon} \, \mathrm{d}x-\omega t +m\theta\right] \right\}$) and the non-oscillatory ($\sim 
|\mathcal{A}(\xi)|^{2} $) terms separately. As a result, to leading order in $x$ it is found
 \begin{equation}\label{eq:psi2sol}
\begin{split}
\mathbf{A_{\Psi{2a}}}(x) & = \begin{pmatrix} 4 \mathcal{D}  x \\[0.1cm] 4x \\[0.1cm] 2ix \end{pmatrix},\\[0.1cm]
\mathbf{A_{\Psi{2b}}}(x) & = \begin{pmatrix} 0 \\[0.1cm] 8x \\[0.1cm] 4 \left(\frac{4}{3} + \gamma \right) \nu_{0} \, x^3 \end{pmatrix}.
\end{split}
\end{equation} 
Again, higher order corrections in $x\ll1$ are omitted here.\\
One notes that the phase shifts between the components of (\ref{eq:psi2sol}) are the same as for the components of (\ref{eq:nullspace}).
With the particular solution (\ref{eq:psi2sol}), the full solution of $\mathit{\hat{L}} \mathbf{\Psi_{2}}  = 
\mathbf{N_{2}}(\mathbf{\Psi_{1}},\mathbf{\Psi_{1}})$ is given by
 \begin{equation}\label{eq:fullpsi2}
\begin{split}
\mathbf{\Psi}_{2} & = \\
\quad & \mathcal{A}(\xi)^{2} \, \mathbf{A_{\Psi{2a}}}(x) \,\exp \left\{2 i\,\left[ \int  \frac{k}{\epsilon} \, \mathrm{d}x-\omega t +m\theta\right] \right\}\\
\quad & + |\mathcal{A}(\xi)|^{2} \, \mathbf{A_{\Psi{2b}}}(x)\\
\quad & + \mathcal{C} \, \mathcal{A}(\xi) \, \mathbf{A_{\Psi{1}}}(x) \, \exp\left\{   \int i\, \frac{k}{\epsilon} \, \mathrm{d}x + im\theta-i\omega t  
\right\}\\
\quad & + c.c.
\end{split}
\end{equation} 
with an arbitrary constant $\mathcal{C}$. The solution contains the second harmonics as well as non-wave contributions for the velocities $u$, $v$.
It turns out that the contribution proportional to $\mathcal{C}$, which contains the null solution (\ref{eq:nullspace}), does not contribute to the amplitude 
equation which will be derived 
in the following section. Therefore we may choose $\mathcal{C}=0$.

\subsection{Third Order Solvability Condition and Amplitude Equation}\label{sec:ampeq}
With the solutions for $\mathbf{\Psi}_{1}$ and $\mathbf{\Psi}_{2}$, we are able to compute the third order inhomogeneity 
$\mathbf{N_{3}}\left(\mathbf{\Psi}_{1},\mathbf{\Psi}_{2}\right)$ (cf. Appendix \ref{sec:nlterms}).
Before proceeding we note that it is not necessary to compute the third order vector of state $\mathbf{\Psi_{3}}$, which would require a complete solution of 
the third order equations. We merely need to evaluate the third order solvability condition 
 \begin{equation}\label{eq:n3}
\langle \mathbf{\Psi_{0}}^{ad}|\mathbf{N_{3}}(\mathbf{\Psi_{1}},\mathbf{\Psi_{2}})  + \partial_{\xi}(\mathit{\hat{M}}\mathbf{\Psi_{1}}) \rangle  = 0,
\end{equation} 
which already yields the desired equation of the form (\ref{eq:landaumodel}) for $\mathcal{A}\left(\xi\right)$, i.e. explicit expressions for the coefficient 
functions $g\left(x\right)$ and $l\left(x\right)$.
The result is
 \begin{equation}\label{eq:landauampfreescaled}
\frac{\mathrm{d} \mathcal{A}}{\mathrm{d}\xi} = \delta_{\nu}^2  \left[g_{r}(x)+i \, g_{i}(x)\right] \mathcal{A} - \left[l_{r}\left(x\right)+i \, l_{i}(x)\right] 
\mathcal{A}|\mathcal{A}|^{2}
\end{equation} 
with real-valued coefficient functions $g_{r}\left(x\right)$, $g_{i}\left(x\right)$, $l_{r}\left(x\right)$ and $l_{i}\left(x\right)$ denoting the real and 
imaginary parts of the coefficients, respectively.
Note that the consistency of the multiple scale expansion requires the condition (\ref{sub:beta2}), which was used to arrive at (\ref{eq:landauampfreescaled}), 
as well as the identity $\mathrm{sgn}\left(\delta_{\nu}^2 \right)|\delta_{\nu}^2|=\delta_{\nu}^2$.
The coefficient functions to leading order in $x$ are given by
\begin{subequations}
 \begin{align}
g_{r}\left(x\right) & = \frac{\left(3 \gamma -2 \right) \nu_{0}}{3 \mathcal{D} \epsilon} x^{2} \equiv \hat{g}_{r} x^2,\label{sub:lincoefr}\\[0.4cm]
g_{i}\left(x\right) & = \frac{\left(3 \gamma -2 \right) \nu_{0}^2}{3 \mathcal{D} \epsilon} x^{4} \equiv \hat{g}_{i} x^4,\label{sub:lincoefi}\\[0.4cm]
l_{r}\left(x\right) & = - \left( \frac{4}{\epsilon}  -  \frac{4\left(589 + 204 \gamma + 9 \gamma^2 \right) \nu_{0}}{81 \mathcal{D} \epsilon} \right) x^{4} 
\equiv \hat{l}_{r} x^4 \label{sub:nonlincoefr},\\[0.4cm]
l_{i}\left(x\right) & = \frac{4 x^3}{\epsilon} \equiv \hat{l}_{i} x^3 \label{sub:nonlincoefi},
\end{align} 
\end{subequations}
where the second equations define the constants $\hat{g}_{r}$, $\hat{g}_{i}$, $\hat{l}_{r}$ and $\hat{l}_{i}$.
Returning to un-scaled units $|\delta_{\nu}|^{2} \partial_{\xi} = \partial_{x}$ and $\mathcal{A}\, |\delta_{\nu}| = \tilde{\mathcal{A}}$, we finally write
 \begin{equation}\label{eq:landauampfree}
\frac{\mathrm{d} \tilde{\mathcal{A}}}{\mathrm{d}x} = \delta_{\nu}^{2} \left[ \hat{g}_{r}x^2+ i \, \hat{g}_{i}x^4\right] \tilde{\mathcal{A}} - \left[\hat{l}_{r} 
x^4+i \, \hat{l}_{i}x^3\right] \tilde{\mathcal{A}}|\tilde{\mathcal{A}}|^{2}.
\end{equation} 
We see that the real part of the linear coefficient in this equation is indeed equivalent to (\ref{eq:grsoll}).
Dropping the tildes and writing for the complex amplitude $\mathcal{A}\left(x\right)\equiv |\mathcal{A}|\left(x\right)\, \exp\left\{i\, 
\Theta\left(x\right)\right\}$, we arrive at
\begin{subequations}
 \begin{align}
\frac{\mathrm{d} |\mathcal{A}|}{\mathrm{d}x} & = \delta_{\nu}^{2} \hat{g}_{r} x^2 \, |\mathcal{A}| - \hat{l}_{r} x^4 
\,|\mathcal{A}|^{3},\label{sub:landauamp}\\[0.4cm]
\frac{\mathrm{d}\Theta}{\mathrm{d}x} & = \delta_{\nu}^{2} \hat{g}_{i} x^4  -\hat{l}_{i} x^3  \, |\mathcal{A}|^{2}.\label{sub:landauphi}
\end{align} 
\end{subequations}
An exact solution of (\ref{sub:landauamp}) is derived in Appendix \ref{sec:ampsol}. It reads in terms of the initial amplitude $|\mathcal{A}_{0}|$
 \begin{equation}\label{eq:landauampsol}
|\mathcal{A}|\left(x\right)  =\frac{ |\mathcal{A}_{0}| \exp\left\{ \frac{1}{3} \delta_{\nu}^{2}\hat{g}_{r}\,  x^{3} \right\}}{\sqrt{ 2 \hat{l}_{r} 
|\mathcal{A}_{0}|^{2} \int_{0}^{x}  t^{4} \exp\left\{ \frac{2}{3} \delta_{\nu}^{2} \hat{g}_{r}  t^{3} \right\} \mathrm{d} t +1}}.
\end{equation} 
The precise behavior of the solution is fairly complicated due to the $x$-dependency of the coefficients.
In the case of a viscously overstable ring ($\delta_{\nu}^2 >0$) the solution (\ref{eq:landauampsol}) first grows exponentially for $x\gtrsim 0$:
 \begin{equation}
|\mathcal{A}|\left(x\right)  \approx |\mathcal{A}_{0}| \exp\left\{ \frac{1}{3} \delta_{\nu}^{2} \hat{g}_{r}  x^{3} \right\}.
\end{equation} 
For $x\gtrsim x^{*}$, where $x^{*}$ denotes a critical distance which depends on the used model parameters, the amplitude closely follows the power law
 \begin{equation}\label{sub:powerlawdamp}
|\mathcal{A}|\left(x\right)  \approx \sqrt{\frac{\delta_{\nu}^{2} \hat{g}_{r}}{\hat{l}_{r}}} \frac{1}{x}.
\end{equation} 
Values of $x^{*}$ lie in the range $10^{-3}\sim 10^{-2}$ for realistic model parameters, as shown below. 

We can relate initial values for the wave amplitude $A_{0}$ to the linear (inviscid) satellite torque $T_{s}$ by using Eq. (30) in \citet{goldreich1979c}
 \begin{equation}\label{eq:lintorque}
 T_{s} = -m  r_{L} \frac{  [4 \, \mathcal{D} \,(\epsilon r_{L} \Omega_{L})^2 A_{0}]^{2}   }{4 G}.
\end{equation} 
This expression is valid to leading order in $x$ and we also used (\ref{eq:nullspace}).
It results from the angular momentum luminosity carried by a long trailing wave with constant amplitude $A_{0}$ in the \emph{linear inviscid theory}. This is 
also the value of the accumulated linear satellite torque,
since in the linear inviscid approximation, the wave transports away all the angular momentum which is excited at the resonance by the satellite. 

\begin{figure}[ht!]
     \centerline{\includegraphics[scale=0.45]{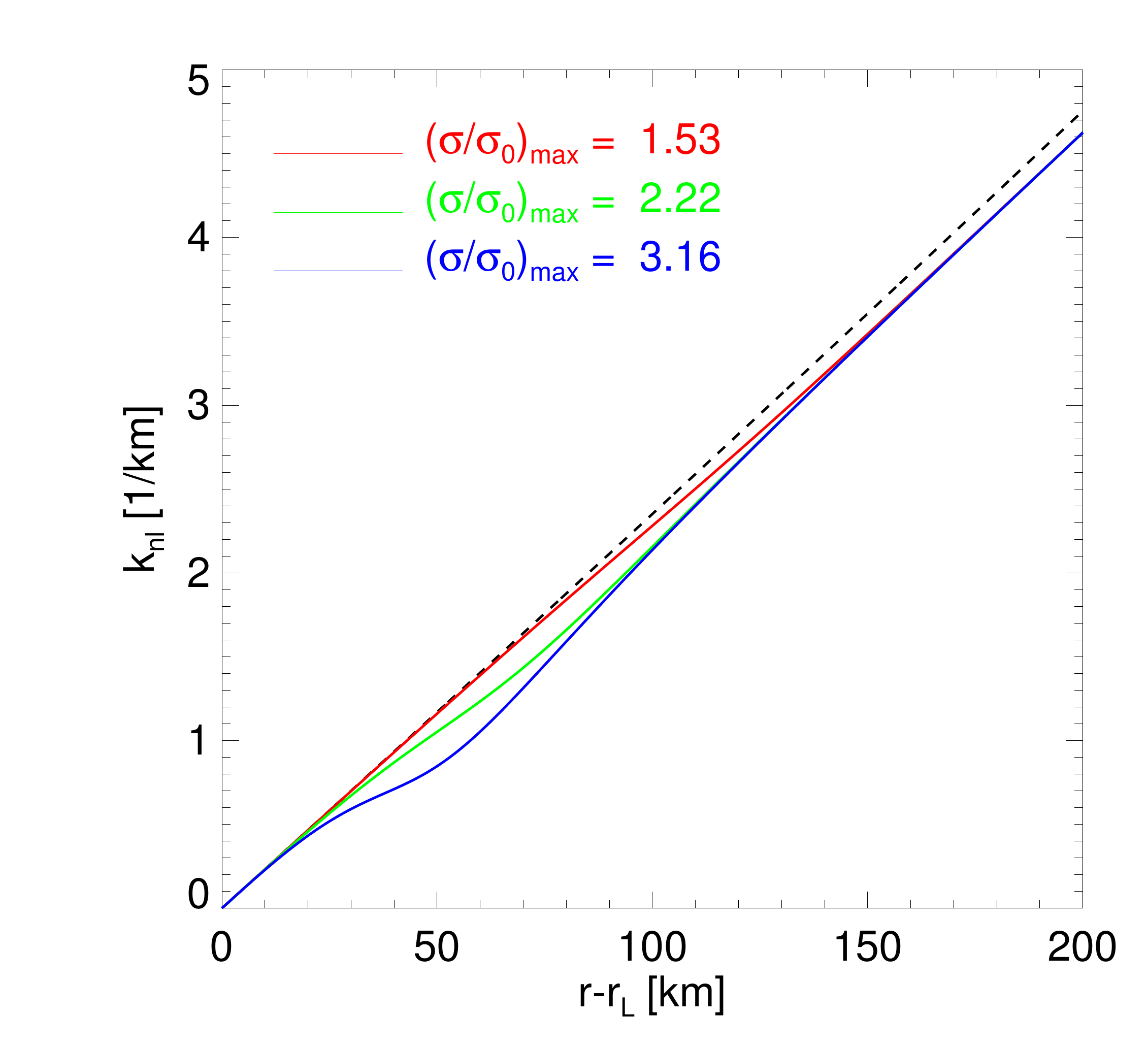}}
     \caption{Nonlinear dispersion relations (\ref{eq:knl}) for density waves with increasing torque values at resonance $T_{s}= -6.7\cdot 10^{8} \, 
\text{kg}\,\text{m}^2 \,\text{s}^{-2}\, (red),\,  -1.1\cdot 10^{10} \, \text{kg}\,\text{m}^2 \, \text{s}^{-2}\, (green),\, -4.3\cdot 10^{10} \, 
\text{kg}\,\text{m}^2 \, \text{s}^{-2} \, (blue)$. 
     We used the \emph{$\tau_{15}$}-parameters (from Table \ref{tab:scalings}) with $r_{L}=10^8 \, \text{m}$, $\sigma_{0}=350 \, \text{kg}\, \text{m}^{-2}$ and 
$m=4$. The torques $T_{s}$ follow from chosen initial amplitudes $A_{0}$ through (\ref{eq:lintorque}). The dashed line is the linear dispersion relation for 
reference. Also indicated are the 
     maximal values of the density contrast for these nonlinear cases.
     The wave assumes these values at those radial distances where the deviations from the linear dispersion relation are highest.}
     	\label{fig:disprel}
     \end{figure}

Further, with (\ref{eq:landauampsol}) we are able to solve for the nonlinear phase shift given by Eq. (\ref{sub:landauphi}):
 \begin{equation}\label{eq:landauphi}
\Theta\left(x\right) = \Theta_{0} - \hat{l}_{i} \int_{0}^{x} t^{3} \, |\mathcal{A}|^{2}\left(t\right) \mathrm{d} t
\end{equation} 
with $\Theta_{0} = \Theta\left(0\right)$. Thus, nonlinearity introduces a phase shift in addition to the radial phase function $\int \frac{k}{\epsilon} \, 
\mathrm{d}x=\frac{x^2}{2 \epsilon}$ of the long trailing density wave [cf. Eq. (\ref{eq:psi1})].
With 
\begin{equation*}
\begin{split}
\mathbf{\Psi_{1}} & = |\mathcal{A}|\left(x\right) \, \mathbf{A_{\Psi{1}}}(x) \, \exp\left\{i  \Theta\left(x\right) +  \int i\, \frac{k}{\epsilon} \mathrm{d}x + 
im\theta-i\omega t  \right\}\\ 
\quad & +c.c.
\end{split}
\end{equation*}
we can define the \emph{nonlinear wavenumber}
 \begin{equation}\label{eq:knl}
k_{nl}=k + \epsilon \,\frac{\mathrm{d} \Theta}{\mathrm{d} x}. 
\end{equation} 
Figure \ref{fig:disprel} shows example plots for $k_{nl}$ for different satellite torque values.
The plots show that local nonlinear effects of self-gravity, expressed through the coefficient $l_{i}\left(x\right)$ (\ref{sub:nonlincoefi}), give rise to a 
reduction of the local wavenumber. 
The waves in this plot are linearly unstable and their amplitudes follow the relation (\ref{sub:powerlawdamp}) for larger distances from resonance (here 
$r-r_{L}\gtrsim 500 \text{km}$). Therefore their wave numbers depart from the linear limit with growing distance from resonance.
     
Let us now consider the final amplitude equation
 \begin{equation}\label{eq:finaleq}
\begin{split}
\frac{\mathrm{d} |\mathcal{A}|}{\mathrm{d} x} & = \frac{\delta_{\nu}^2 \left(3 \gamma -2 \right) \nu_{0}}{3 \mathcal{D} \epsilon} x^{2} \,|\mathcal{A}|\\[0.4cm]
\quad & + \left( \frac{4}{\epsilon}  -  \frac{4\left(589 + 204 \gamma + 9 \gamma^2 \right) \nu_{0}}{81 \mathcal{D} \epsilon} \right) x^{4} \, |\mathcal{A}|^{3}.
\end{split}
\end{equation} 
For a \textit{viscously overstable ring}, the linear instability of the density wave manifests through the first term with $\delta_{\nu}^2 > 0$, 
corresponding to $\beta > \beta_{c}$ [cf. (\ref{eq:expar})]. 
The case of ``linear viscous damping'' is described by values $\delta_{\nu}^2 < 0$, corresponding to $\beta < \beta_{c}$. Nonlinear damping, described by the 
cubic term $\propto |\mathcal{A}|^{3}$, is dominated physically by viscous terms as well. The first term in the bracket is purely self-gravitational and 
positive. 
Hence, in the limit of small viscosities, this term would, theoretically, cause a nonlinear instability for large values\footnote{which would require the 
derivation of a stabilizing quintic term in the amplitude equation.} of $x$. 
However, for realistic values of the viscosity the self-gravity term is negligible compared to the viscous contribution, which has negative sign.
Note that in Eq. (\ref{eq:finaleq}) the relative magnitude of the nonlinear term $\propto |\mathcal{A}|^{3}$ (strongly) depends on the distance to the Lindblad 
resonance ($\propto x^4 $), and not only on the magnitude of $\mathcal{A}$ itself. This $x$-dependency, and also those of the other coefficients arise from the 
radial dependence of the scaled wavenumber $k(x)= x$.

In conclusion, Eq. (\ref{eq:finaleq}) describes the damping of a density wave under the influence of density dependent viscosities in the weakly nonlinear 
regime. It is a generalization of the linear viscous damping relation with constant viscosities (\ref{eq:gt78}). 
A consequence of Eq. (\ref{sub:powerlawdamp}) is that if the condition for viscous overstability ($\delta_{\nu}^2 >0$) is fulfilled, the surface density 
perturbation of the weakly nonlinear model does not decay to zero but rather saturates to a finite value at large distance from the resonance. Namely, the 
WKB-solution (\ref{eq:poisson1}) of Poisson's equation and the first component of the null vector (\ref{eq:nullspace}) yield the following expression for the 
first order density perturbation:
 \begin{equation}\label{eq:sigma1full}
\sigma_{1} = -4x \,|\mathcal{A}|\left(x\right) \sin\left(\frac{x^2}{2 \epsilon}+\Theta\left(x\right) + m\theta - \omega t\right).
\end{equation} 
With the asymptotic solution for $|\mathcal{A}|$ for large $x$ given by (\ref{sub:powerlawdamp}), it follows that the amplitude of the unscaled surface density 
perturbation $\sigma\left(x\right)$ saturates to a constant value
 \begin{equation}\label{eq:sigmasat}
\sigma\left(x \to \infty\right) \approx 4 \sigma_{0} \sqrt{\delta_{\nu}^2 \, \hat{g}_{r}/\hat{l}_{r}}.
\end{equation} 
The saturation to this value occurs for smaller $x$ if $\beta$ increases. 

\begin{figure}[ht!]
     \centerline{\includegraphics[scale=0.45]{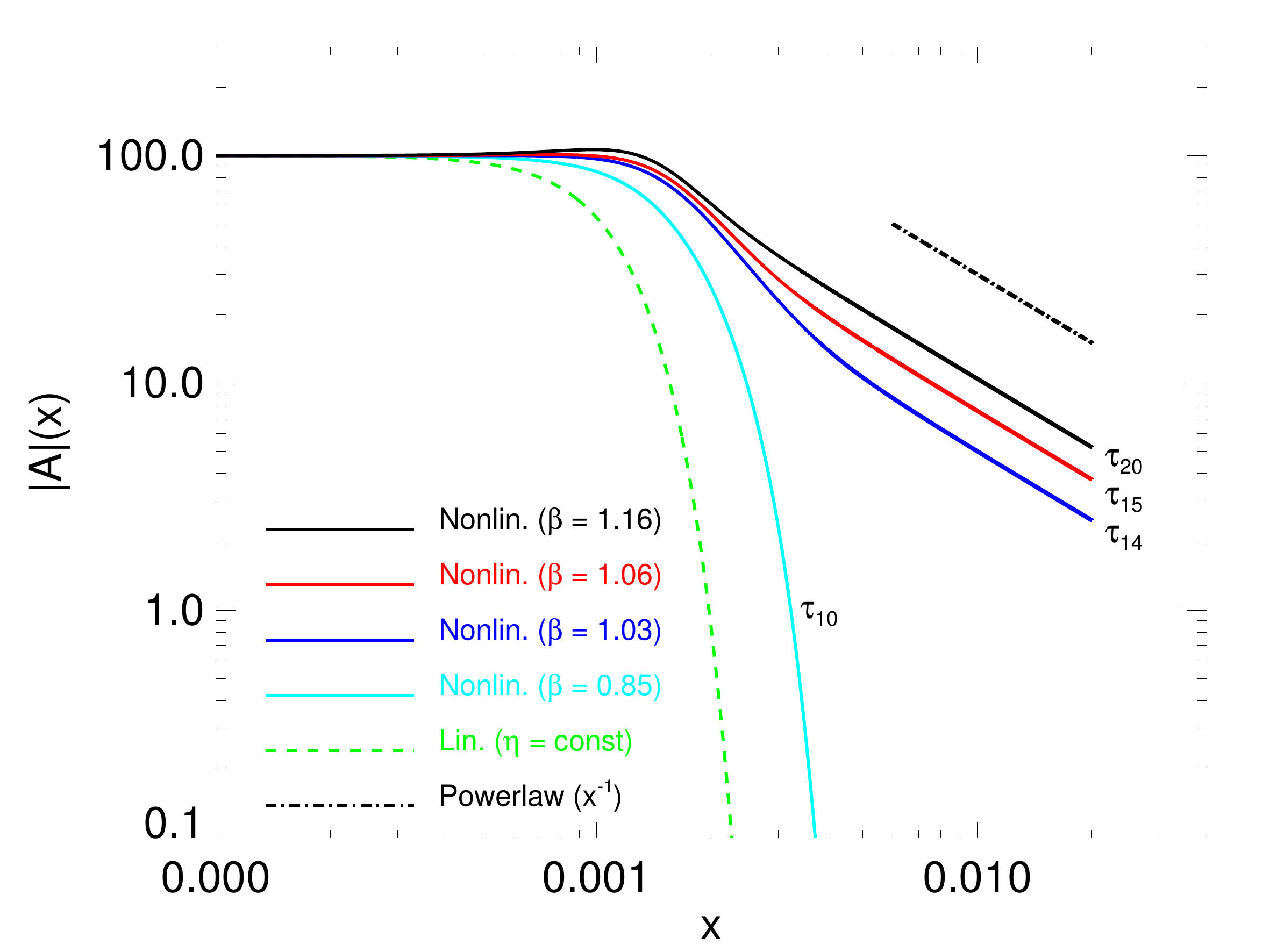}}
     \caption{Numerical solutions of the nonlinear amplitude equation (\ref{eq:finaleq}) for the parameter sets listed in Table \ref{tab:scalings} with $m=4$. 
Further, we used $r_{L}=10^8\, \text{m}$ and $\sigma_{0}=350\,\text{kg}\, \text{m}^{-2}$. The green dashed line shows the linear viscous damping relation in 
the 
limit of constant viscosity ($\nu_{0}$ and $\gamma$ from the \emph{$\tau_{20}$}-parameter set). Note that for the \emph{$\tau_{10}$}-parameters the condition 
for viscous overstability is not fulfilled and eventually the amplitude damps exponentially. The initial amplitude $|\mathcal{A}_{0}|=100$ at $x=0$ corresponds 
to a satellite torque $T_{s}=-9.54\cdot 10^{8}\, \text{kg}\,\text{m}^2 \,\text{s}^{-2}$.}
     	\label{fig:ampsol}
     \end{figure}

In Figure \ref{fig:ampsol} numerical solutions of the amplitude equation (\ref{eq:finaleq}) are plotted for different parameter sets.
These parameter sets were obtained in N-body simulations [\citet{salo2001}, \citet{schmidt2003}] and are listed in Table \ref{tab:scalings}.
One notes that the values of $\delta_{\nu}$ (also provided in Table \ref{tab:scalings}) corresponding to these parameters are
significant fractions of unity. Thus, we expect our model to be qualitatively correct for these parameters, though not quantitatively.
Turning to the discussion of Figure \ref{fig:ampsol}, the dashed green curve represents the linear viscous damping relation, which results from the limit of a 
vanishing density dependence of viscosities. 
This is an unrealistic assumption for dense planetary rings, as outlined in Section \ref{sec:lintheo}.
In the nonlinear cases, the amplitude decays substantially slower. The N-body parameter sets with optical depths $\tau=1.4$, $\tau=1.5$ and $\tau=2$ and 
corresponding viscous parameters $\beta=1.03$, $\beta=1.06$ and $\beta=1.16$, fulfill 
the (hydrodynamic) condition for viscous overstability. For these cases we observe a turnover to a power law damping relation (\ref{sub:powerlawdamp}). 
This turnover occurs for smaller $x$, the larger the value of $\beta$, or, equivalently, of $\tau$.
The parameter set with $\tau=1.0$ (and $\beta=0.85$) does not exhibit overstability, since $\beta < \beta_{c}$. 
For this linearly stable case, the damping behavior follows the nonlinear solutions with $\beta >\beta_{c}$ for a certain range, but eventually turns into an 
exponential decay. The reason for this nonlinear behavior for small $x$ is that the initial amplitude $|\mathcal{A}_{0}|$ is chosen fairly high, such that the 
wave becomes already nonlinear within a few wavelengths from resonance.
     
The amplitude $|\mathcal{A}|$ is related to the nonlinearity parameter $q$ of a streamline model by 
 \begin{equation}\label{eq:qpar}
 q=4\, x |\mathcal{A}|  
\end{equation} 
which follows from equations (7) and (8a) in \citet{longaretti1986} and Eq. (\ref{eq:sigma1full}) of this paper. 
In the streamline model, perturbed ring matter is described in terms of eccentric streamlines and $q$ measures the
radial displacement of adjacent streamlines, relative to their unperturbed distance. Generally we have $0\leq q < 1$.
The unperturbed state corresponds to $q=0$, whereas in a strongly nonlinear wave one finds $q\lesssim 1$.
For a detailed description of the streamline formalism we refer to the papers listed in Section \ref{sec:intro}.

\begin{figure}[ht!]
     \centerline{\includegraphics[scale=0.45]{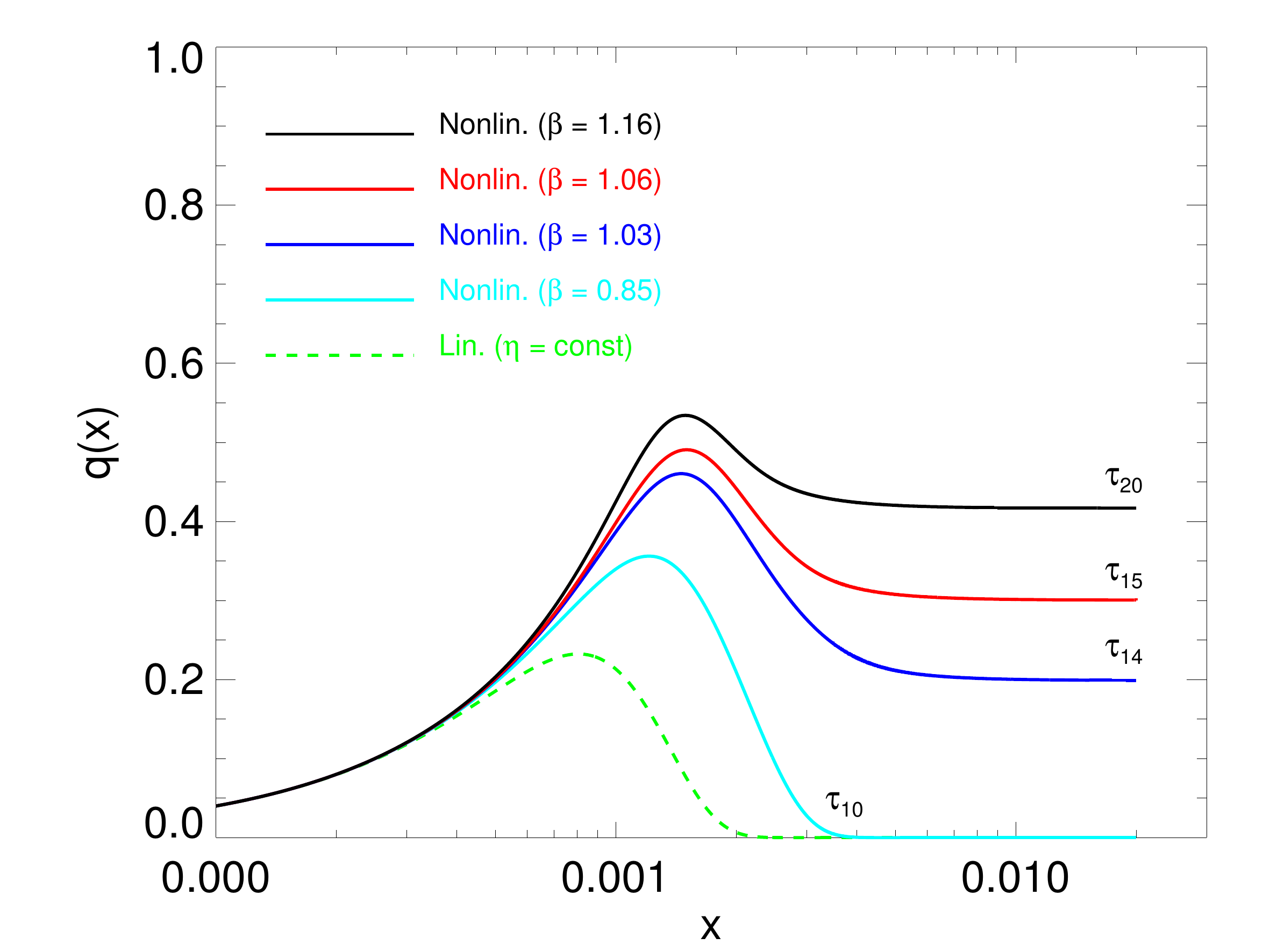}}
     \caption{Radial profiles of the nonlinearity parameter $q$ corresponding to the amplitude solutions in Fig. \ref{fig:ampsol}.}
     	\label{fig:ampsolqq}
     \end{figure}

\begin{figure*} [ht!]
     \centerline{\includegraphics[width=0.9 \textwidth]{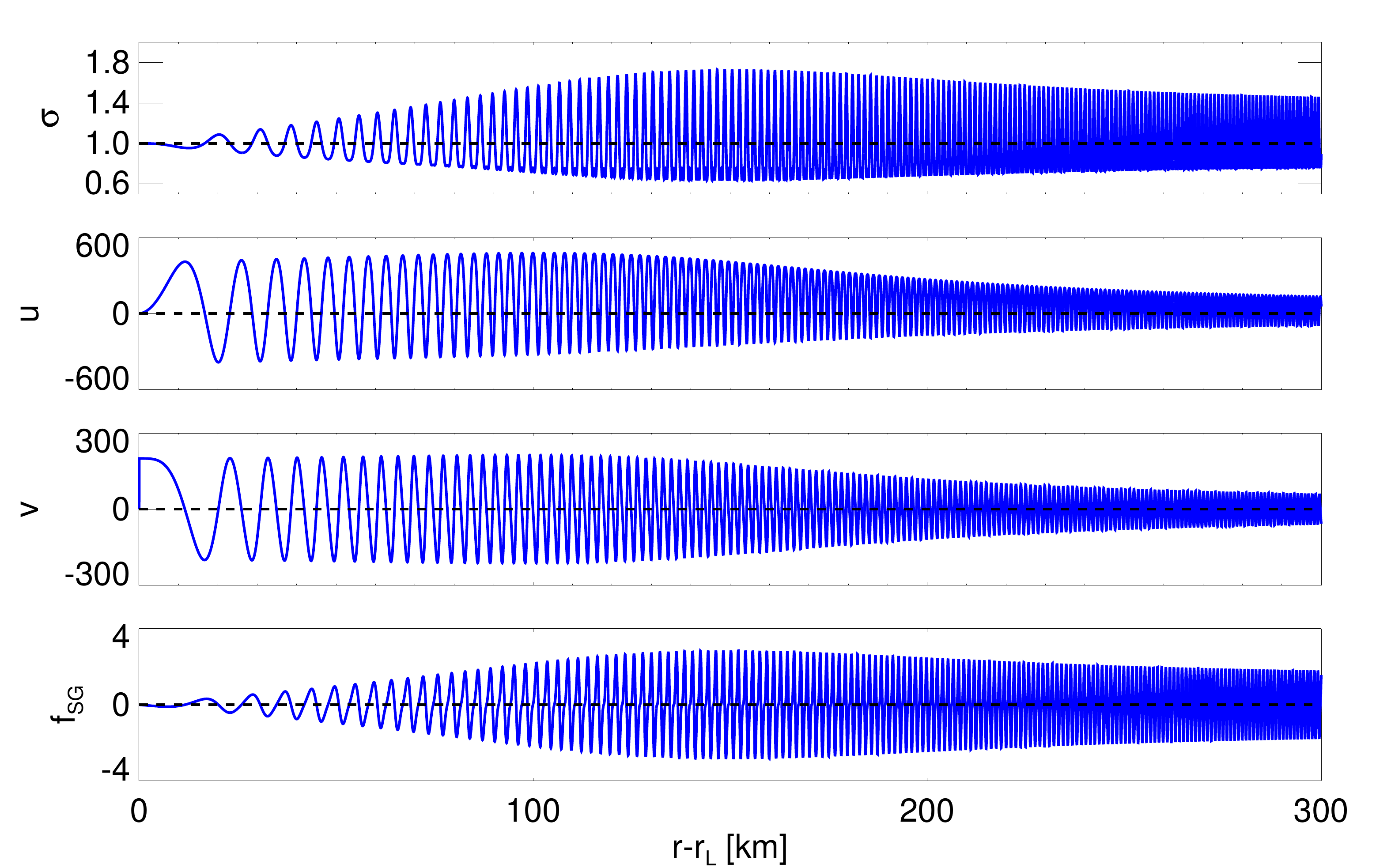}}
     \caption{Scaled hydrodynamic quantities to order $|\delta_{\nu}|^2$ related to the density wave corresponding to the 
     case ($\tau=1.5;~\beta=1.06$) in Figure \ref{fig:ampsol}. 
     Also shown is $f_{SG}$, the scaled self-gravity force, defined in (\ref{eq:sgforce}).}.
     	\label{fig:fieldsol}
\end{figure*}

Relation (\ref{eq:qpar}) is, strictly speaking, valid only in the
weakly nonlinear limit. In this limit $q$ is equal to the amplitude of the first order density perturbation (\ref{eq:sigma1full}). 
In Figure \ref{fig:ampsolqq} we present the radial profiles of $q$ corresponding to the amplitude solutions in Figure \ref{fig:ampsol}.
One notes a saturation of $q$ for those parameters that fulfill the condition for viscous overstability.
Further, Fig. \ref{fig:fieldsol} shows the hydrodynamic field quantities corresponding to the case $\beta=1.06$ in Figure \ref{fig:ampsol}. 
The surface density oscillations persist indefinitely. In the same figure, $f_{SG}$ is the scaled self-gravity force per unit mass
 \begin{equation}\label{eq:sgforce}
\begin{split}
 f_{SG}\left(x\right)& = -\frac{\partial \phi}{\partial r}\\[0.1cm]
&  \quad  = i \frac{\mathcal{D}}{ \epsilon} \, \sigma\left(x\right) + c.c. 
 \end{split}
\end{equation} 
where we used the solution of the Poisson Eq. (\ref{eq:poissonsol}). Recall that $\mathcal{D}$ is scaled with $\Omega_{L}^2$.
The plots in Figure \ref{fig:fieldsol} are the \emph{second order representation of the vector of state} (\ref{eq:msexpvec}), 
where we used (\ref{eq:psi1}), (\ref{eq:marginalk}), (\ref{eq:nullspace}), (\ref{eq:psi2sol}) and (\ref{eq:fullpsi2}). 
As a result, the density profiles are not entirely smooth, particularly in the density minima, lacking corrections by the orders $>2$. 
Nevertheless, the second order contribution in (\ref{eq:msexpvec}) correctly leads to a flattening of the density minima and a sharpening of the maxima. 
This is in contrast to the density wave profiles that result from `streamline models' which are based on a Lagrangian equation of continuity.
However, the wave amplitude which is computed from Eq. (\ref{eq:landauampfree}), is not affected by the restriction on second order harmonics.
In Section \ref{sec:bgtcomp} we compare our weakly nonlinear model with the streamline model applied in \citet{bgt1986}. 

\subsection{Non-WKB Effects of Self-Gravity}\label{sec:wkbsg}

The model derived in the previous sections includes self-gravity effects that go beyond the WKB-approximation.
In Appendix \ref{sec:poisson} we solve the Poisson equation while including the effects of the slow length scale, i.e. the slow change of amplitude. 
The solution (\ref{sub:poisson3}) includes the term
 \begin{equation}\label{eq:sgcorr}
 \frac{i  \mathrm{s} \epsilon }{\mathcal{D} }  \frac{\partial \phi_{1}}{\partial \xi}
\end{equation} 
which describes the slow amplitude-related change of the self-gravity potential (in the lowest approximation). 
\begin{figure*}[h!] 
     \centerline{\includegraphics[width= \textwidth]{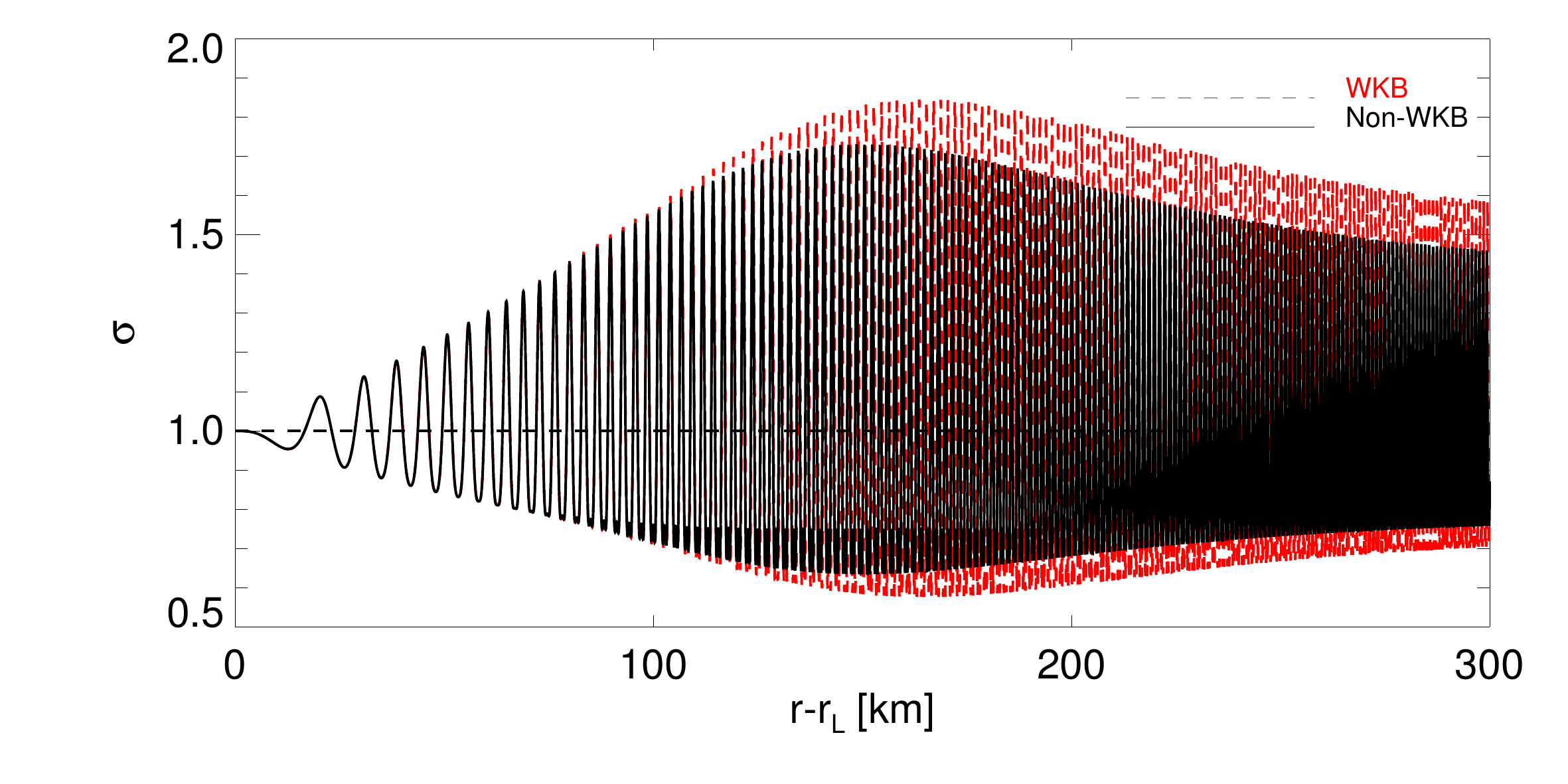}}
     \caption{Illustration of the self-gravity correction due to amplitude change on the density wave profile. The parameters are the 
$\tau_{15}$-parameters with $m=4$ and $T_{s}= -9.54\cdot 10^{8} \, \text{kg}\,\text{m}^2 \,\text{s}^{-2}$. The wave profile labeled ``Non-WKB'' has been 
computed by using the coefficient functions (\ref{sub:lincoefr}-d).
     For the profile labeled ``WKB'' we re-derived the coefficient functions from a multiple scale expansion \emph{without} the corrective term 
(\ref{eq:sgcorr}). Further, we used $r_{L}=10^8 \, \text{m}$ and $\sigma_{0}=350\, \text{kg} \, \text{m}^{-2}$. Note that small cusps in the density minima 
result from the omission of higher-order terms in our theory (see the discussion at the end of Section \ref{sec:ampeq}).}
     	\label{fig:nonwkb}
\end{figure*}
In contrast, the WKB-result involves only derivatives of the rapidly varying phase
of the potential. We generally observe that this correction to the WKB-order gives rise to shorter damping lengths of density wave profiles. An example is shown 
in Figure \ref{fig:nonwkb}.

\newpage

\section{Nonlinear Damping of Forced Density Waves}\label{sec:nlforced}

In this section we include the forcing by an external satellite in our analysis of Eqs. (\ref{eq:nleq}) and derive a nonlinear amplitude equation describing the 
propagation of \emph{forced} density waves subject to viscous stress. 

The wave is excited by one particular Fourier mode of the potential of an orbiting satellite.
We neglect orbital eccentricity and inclination of the satellite. Thus, we consider only first order resonances of the type $m:m-1$.
This restriction is made for the sake of simplicity of the calculations and does not occlude any physical aspects that are investigated here.
We adopt again the scalings of quantities and parameters as given in Table (\ref{tab:scalings}). 

We can apply the multiple scale analysis presented in Section \ref{sec:nlfree} to resonantly forced density waves with slight modifications in the derivation. 
The \textit{scaled} resonant mode of the forcing satellite potential reads
 \begin{equation}
\begin{split}
\phi_{s}\left(x,\theta,t\right)  & = -\frac{G M_{s}}{a \,\epsilon^2 r_{L}^2 \Omega_{L}^2} \,b_{12}^{m}\left(x\right) \exp\left(i\, m \theta - i\, \omega 
t\right)\\
&  \quad  + c.c.
\end{split}
\end{equation} 
where $a$ is the satellite's semi major axis and $M_{s}$ its mass. Further, $b_{12}^{m}\left(x\right)$ is a Laplace-coefficient:
\begin{equation}
b_{12}^{m}\left(x\right)   = \frac{2}{\pi} \int_{0}^{ \pi} \mathrm{d}\Psi \frac{\cos\left(m \Psi\right)}{\sqrt{1+ \rho\left(x\right)^2 - 2 \rho \left(x\right) 
\cos\left( \Psi\right)}}
\end{equation} 
with
 \begin{equation}
\rho = \frac{r}{a}\, , \hspace{0.1 cm}\text{and} \hspace{0.15 cm} r  = r_{L}\left(1+x\right).
\end{equation} 
Note that we evaluate the satellite forcing terms in Eqs. (\ref{eq:nleq}) at the resonance location with $\rho\left(x=0\right) =r_{L}/a=[(m-1)/m]^{2/3}$ 
(\citet{goldreich1978b}).\\
The definition of the expansion parameter now reads
 \begin{equation}\label{eq:deltafor}
\delta_{s} = \left[\frac{G M_{s}}{a\, \epsilon r_{L}^2 \Omega_{L}^2}\right]^{1/3} 
\end{equation} 
which describes the ratio of satellite forcing to the self-gravity force. 
This parameter is very similar to the forcing parameter $f$ in Eq. (36) in \citet{shu1985a}.
Strictly, it must have values much smaller than unity to ensure the validity of the weakly nonlinear analysis.
Using the values $r_{L}=96248\,\text{km}$ (\citet{hedman2016}), $M_{S}=1.898 \cdot 10^{18}\, \text{kg}$ (\citet{jacobson2008}) and $m=2$, corresponding to the 
(strongly nonlinear) 
Janus 2:1 density wave, as well as $\sigma_{0}=600\, \text{kg}\,\text{m}^{-2}$, one obtains $\delta_{s}= 0.47$.
The linear (inviscid) satellite torque value corresponding to this forcing strength is $T_{S}= -3.61\cdot 10^{11} \, \text{kg}\,\text{m}^2\, \text{s}^{-2}$, 
obtained with Eq. (\ref{eq:lintor}) below.
For such large value of $\delta_s$ we do not expect our model to 
produce \emph{quantitatively} correct results.
For illustrative examples which follow below we will use the above values for $r_{L}$ and $\sigma_{0}$ in combination with various viscosity parameters and 
different values of $\delta_{s}$.

We now expand the vector of state in powers of the parameter $\delta_{s}$ as we did in Eqs. (\ref{eq:msexpvec}), (\ref{eq:msexpvec2}) with the parameter 
$\delta_{\nu}$.
Since $\delta_{s} >0$ always, we do not need to work with its absolute value.
The forcing terms will appear now at $\mathcal{O}\left(\delta_{s}^3\right)$ of the expansion. This is desirable since we want to obtain an equation describing 
the damping
of forced waves, as we already know that nonlinear viscous damping occurs at this order. 
We again expand the viscous parameter [cf. (\ref{eq:shearvis}), (\ref{sub:betaexp})]
 \begin{equation}\label{eq:betaexps}
\beta = \beta_{c} + \delta_{s} \, \beta_{1} + \delta_{s}^2 \, \beta_{2} +\cdots \, .
\end{equation} 
As in Section \ref{sec:secondorder}, the second order solvability condition yields $\beta_{1}=0$.
However, important to note is that in the current situation $\beta_{2}$ is not restricted (by the definition of the expansion parameter $\delta_{s}$) to yield 
$\beta_{2}=\beta_{c}$,
as it was the case in Section \ref{sec:msexp}.
But the contribution $\delta_{s}^2 \, \beta_{2}$ should strictly be much smaller than $\beta_{c}$ for the consistency of the expansion 
(\ref{eq:betaexps}). 
We can define $\beta_{2}$ such that
 \begin{equation}
\delta_{s}^2 \, \beta_{2} \equiv \delta_{\nu} ^2 \, \beta_{c}
\end{equation} 
where the free parameter $\delta_{\nu}$ is identical to (\ref{eq:expar}) and controls the distance of the system to the threshold for viscous 
overstability.
With these definitions the derivations of the vector of state components at $\mathcal{O}\left(\delta_{s}\right)$ and $\mathcal{O}\left(\delta_{s}^2\right)$ 
yield 
the same results as in Sections \ref{sec:linstab} and \ref{sec:secondorder}.
However, at $\mathcal{O}\left(\delta_{s}^3\right)$, we obtain a modified inhomogeneity
\begin{equation}\label{eq:inhom3f}
  \mathbf{N_{3}^{f}} = \mathbf{N_{3}}\left(\mathbf{\Psi}_{1},\mathbf{\Psi}_{2}\right) + \mathbf{f}
\end{equation}
where
\begin{equation}
 \mathbf{f}=\begin{pmatrix} 0  \\[0.1cm] \left[\partial_{x} b_{12}^{m}(x)\right]_{r_{L}}\, \cos \left(m \theta - \omega t \right) \\[0.1cm] -m\, 
\left[b_{12}^{m}(x)\right]_{r_{L}}\, \sin\left(m \theta - \omega t\right)  \end{pmatrix}
\end{equation}
describes the satellite forcing.
From the inhomogeneity (\ref{eq:inhom3f}) we arrive at a modified third order solvability condition, which now includes the additional term
 \begin{equation}\label{eq:forcing}
 \mathbf{F}(x)  = \frac{1}{2\pi} \int_{0}^{2 \pi} \mathrm{d} \theta \,\exp\left\{-\int i\, \frac{k}{\epsilon} \mathrm{d}x + im\theta-i\omega t  \right\} \, 
 \langle  \mathbf{A_{\Psi{0}}}^{ad}(x) |  \mathbf{f}\rangle.
\end{equation} 
with the adjoint null vector $\mathbf{A_{\Psi{0}}}^{ad}(x)$ given by (\ref{eq:nullspacead}) and where the scalar product $\langle \cdots |\cdots \rangle$ 
is defined through (\ref{eq:scalarprod}).
Thus, compared to the free wave analysis in Section \ref{sec:nlfree}, the nonlinear amplitude equation is changed according to $\frac{\mathrm{d} 
\mathcal{A}}{\mathrm{d} \xi} \to \frac{\mathrm{d} \mathcal{A}}{\mathrm{d} \xi} - \mathbf{F}(x)$.
Evaluating the integral over $\theta$ in (\ref{eq:forcing}) yields
\begin{equation}
\begin{split}
\mathbf{F}(x) & = \frac{ \left[ \omega\, \partial_{x} b_{12}^{m}(x)- \, m \Omega\left( 2\, b_{12}^{m}(x) + \partial_{x} b_{12}^{m}(x) \right) \right]_{r_{L}}}{4 
i \Omega }\\
&  \quad  \cdot \exp\left\{-\int i\, \frac{k}{\epsilon} \mathrm{d}x \right\}.
\end{split}
\end{equation} 
Returning to ordinary length scale $\partial_{\xi} \to (1/\delta_{s}^2) \partial_{x}$, $\mathcal{A} \to (1/\delta_{s}) \tilde{\mathcal{A}}$, 
dropping the tildes and using the relations (\ref{eq:dexp}), (\ref{eq:marginalk}), (\ref{sub:freq1}) and (\ref{sub:freq2}) for a long trailing density wave 
in the vicinity of an ILR, the nonlinear amplitude equation now reads
 \begin{equation}\label{eq:fampeq}
\begin{split}
\frac{\mathrm{d} \mathcal{A}}{\mathrm{d} x} & =  \left[f_{r}(x)+i f_{i}(x)\right] + \delta_{\nu}^2 \left[g_{r}(x)+i g_{i}(x)\right] \,\mathcal{A}\\[0.1cm]
&  \quad  -\left[l_{r}(x)+i l_{i}(x)\right] \,|\mathcal{A}|^2 \mathcal{A},
\end{split}
\end{equation} 
where the leading orders of the coefficient functions are given by
\begin{subequations}
 \begin{align}
f_{r}\left(x\right) & = -\frac{\partial_{x} \phi_{s} \left(r_{L} \right)+2m\, \phi_{s}\left(r_{L}\right) }{4 \mathcal{D}} \sin\left(\frac{x^2}{2 
\epsilon}\right)\label{sub:forcecoefr},\\[0.2cm]
f_{i}\left(x\right) & = -\frac{\partial_{x} \phi_{s}\left(r_{L}\right)+2m \,\phi_{s}\left(r_{L}\right)}{4 \mathcal{D}} \cos\left(\frac{x^2}{2 
\epsilon}\right)\label{sub:forceoefi},\\[0.2cm]
g_{r}\left(x\right) & = \frac{\left(3 \gamma -2 \right) \nu_{0}}{3 \mathcal{D} \epsilon} x^{2},\label{sub:lincoefr2}\\[0.2cm]
g_{i}\left(x\right) & = \frac{\left(3 \gamma -2 \right) \nu_{0}^2}{3 \mathcal{D} \epsilon} x^{4},\label{sub:lincoefi2}\\[0.2cm]
l_{r}\left(x\right) & = - \left( \frac{4}{\epsilon}  -  \frac{4\left(589 + 204 \gamma + 9 \gamma^2 \right) \nu_{0}}{81 \mathcal{D} \epsilon} \right) 
x^{4}\label{sub:nonlincoefr2},\\[0.2cm]
l_{i}\left(x\right) & = \frac{4 x^3}{\epsilon}\label{sub:nonlincoefi2}.
\end{align} 
\end{subequations}
Clearly, the functions $g_{r}$, $g_{i}$, $l_{r}$ and $l_{i}$ are equivalent to (\ref{sub:lincoefr})-(\ref{sub:nonlincoefi}).
For convenience we inserted the wavenumber $k=x$  and solved the phase integral $\exp\left\{-\int i\, \frac{k}{\epsilon} \mathrm{d}x \right\}$ while assuming 
zero value at the resonance. 
The magnitude $|\mathcal{A}|\left(x\right)$ is defined as $|\mathcal{A}|\left(x\right) = \left[ \mathcal{A}_{r}\left(x\right)^2 + 
\mathcal{A}_{i}\left(x\right)^2 \right]^{1/2}$,
where the terms on the right hand side denote the real and imaginary parts of $\mathcal{A}(x)$.
If we split Eq. (\ref{eq:fampeq}) into its real and imaginary parts, we obtain the two equations
\begin{subequations}
 \begin{align}
\begin{split}
\frac{\mathrm{d} \mathcal{A}_{r}}{\mathrm{d} x}  &= f_{r}(x) + \delta_{\nu}^2 \left[g_{r}(x) \,\mathcal{A}_{r} -g_{i}(x) \,\mathcal{A}_{i}  \right] 
\label{sub:fampeq1}\\
&  \quad  - |\mathcal{A}|^2 \, \left[l_{r}(x) \, \mathcal{A}_{r}-l_{i}(x) \, \mathcal{A}_{i} \right],
\end{split}
\end{align} 
 \begin{align}
\begin{split}
\frac{\mathrm{d} \mathcal{A}_{i}}{\mathrm{d} x}  & = f_{i}(x) + \delta_{\nu}^2 \left[g_{i}(x) \,\mathcal{A}_{r} +g_{r}(x) \,\mathcal{A}_{i}  \right] 
\label{sub:fampeq2}\\
&  \quad  - |\mathcal{A}|^2 \, \left[l_{r}(x) \, \mathcal{A}_{i}+l_{i}(x) \, \mathcal{A}_{r} \right].
\end{split}
\end{align} 
\end{subequations}
These coupled equations have to be solved simultaneously under the restriction of a given boundary condition. From the linear theory of satellite forcing 
(\citet{goldreich1979c}; \citet{shu1984}) we know that a suitable condition is given by
$\mathcal{A}(x\to -\infty) = 0$. 

\begin{figure*} [ht!]
     \centerline{\includegraphics[width=0.6 \textwidth]{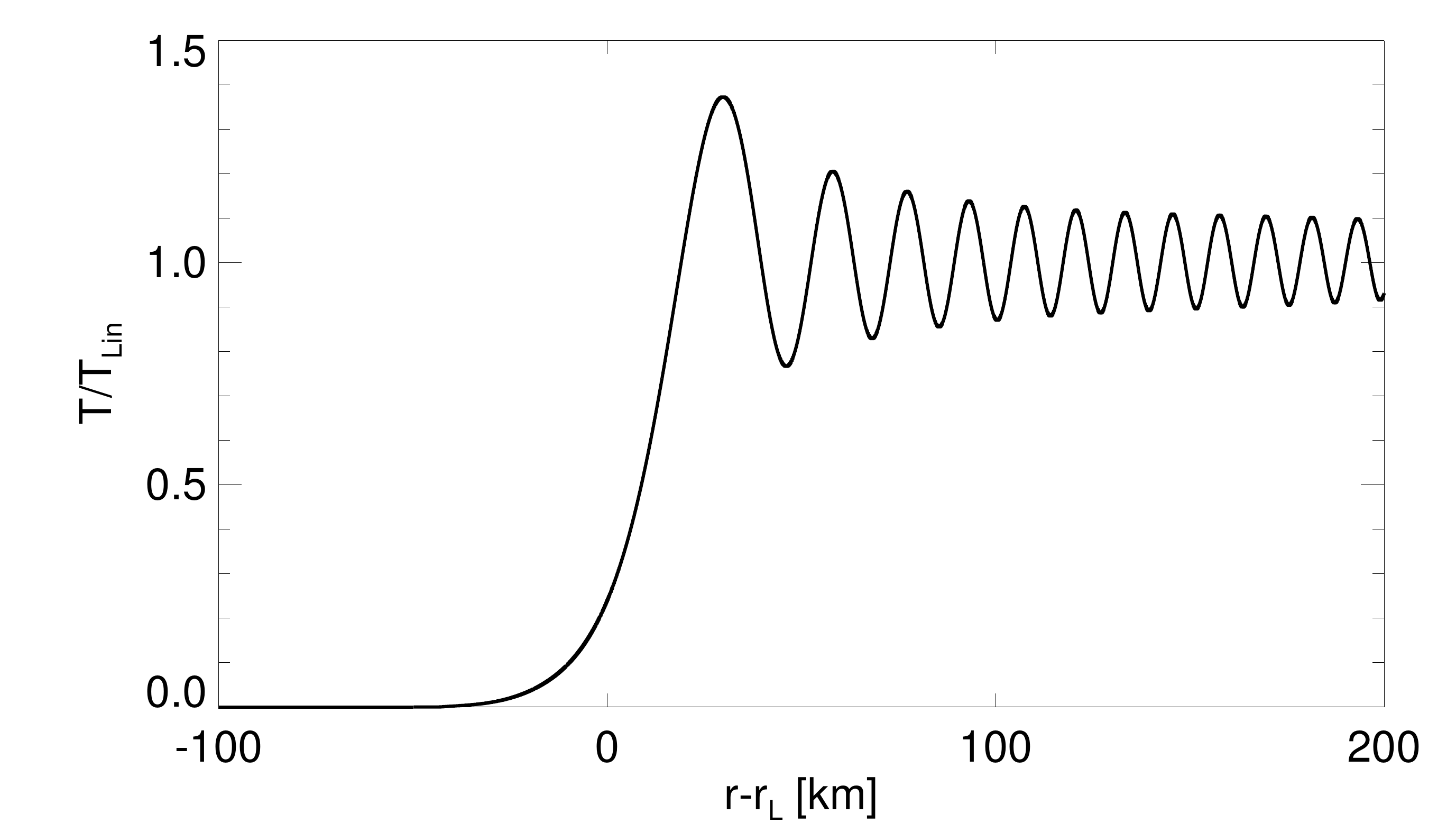}}
     \caption{The scaled accumulated torque following from numerically solving Eqs. (\ref{eq:tordens})-(\ref{eq:lintor}) for a $m=2$ density wave with the 
$\tau_{15}$-parameters and $\delta_{s} = 0.37$. Further, we used $r_{L}=96248\,\text{km}$ and $\sigma_{0}=600\, \text{kg}\,\text{m}^{-2}$.}
     	\label{fig:torque}
\end{figure*}

With equation (\ref{eq:fampeq}) we are now able to describe the \emph{excitation} of the nonlinear density waves whose propagation and damping behavior we 
derived in Section \ref{sec:nlfree}.
An expression for the (weakly) nonlinear satellite torque can be derived as follows. Recall that the angular momentum luminosity carried by a free wave with 
amplitude $\mathcal{A}$ is given through [cf. Eq. (\ref{eq:lintorque})]
 \begin{equation}
 L\left(x\right) = -m  r_{L} \frac{  \left[4  \mathcal{D} \,(\epsilon r_{L} \Omega_{L})^2 |\mathcal{A}|\right]^{2}   }{4 G}.
\end{equation} 
Let us consider the radial derivative of $L$ by using (\ref{eq:fampeq}) 
 \begin{equation}\label{eq:lflux}
\begin{split}
\frac{\mathrm{d}L}{\mathrm{d}x} & = - m  r_{L}\frac{  \left[4  \mathcal{D} \,(\epsilon r_{L} \Omega_{L})^2 \right]^{2}}{4 G} \,\frac{\mathrm{d} 
|\mathcal{A}|^{2}}{\mathrm{d} x}\\[0.1cm]
\quad & = - m  r_{L}\frac{ \left[4  \mathcal{D} \,(\epsilon r_{L} \Omega_{L})^2 \right]^{2}}{4 G} \left( \frac{\mathrm{d} \mathcal{A}}{\mathrm{d} x} \cdot 
\mathcal{A}^{*} + \frac{\mathrm{d} \mathcal{A}^{*}}{\mathrm{d} x} \cdot \mathcal{A}\right)\\[0.1cm]
\quad & = - m  r_{L} \frac{  \left[4  \mathcal{D} \,(\epsilon r_{L} \Omega_{L})^2 \right]^{2}}{2 G} \left(f_{r}\left(x\right)\mathcal{A}_{r} 
+f_{i}\left(x\right)\mathcal{A}_{i}  +  \delta_{\nu}^2 g_{r}\left(x\right) |\mathcal{A}|^2 -  l_{r}\left(x\right) |\mathcal{A}|^4 \right),
\end{split}
\end{equation} 
where a star denotes complex conjugate. From this equation we can identify the satellite torque density
 \begin{equation}\label{eq:tordens}
\mathcal{T}=- m  r_{L} \frac{  \left[4  \mathcal{D} \,(\epsilon r_{L} \Omega_{L})^2 \right]^{2}   }{2 G} \left(f_{r}\left(x\right)\mathcal{A}_{r} 
+f_{i}\left(x\right)\mathcal{A}_{i} \right).
\end{equation} 
The remaining terms in (\ref{eq:lflux}) describe viscous effects on the angular momentum luminosity of the wave. 
The torque density (\ref{eq:tordens}) results in an accumulated torque function
 \begin{equation}\label{eq:acctor}
T\left(x\right) = \int_{-\infty}^{x} \mathcal{T}\left(\hat{x}\right) \mathrm{d} \hat{x}
\end{equation} 
which denotes the accumulated satellite torque at radial location $x$.
For all parameter regimes considered in this paper we find that the torque values $T \left(x\right)$ for large $x$ are nearly equal to the linear inviscid 
values (\citet{goldreich1979c})
 \begin{equation}\label{eq:lintor}
T_{Lin}= -m \pi^2 \frac{ \sigma_{0}}{\mathcal{D} \, \Omega_{L}^2} \left(\epsilon \, r_{L} \Omega_{L}\right)^4 \left[ \partial_{x}\phi_{s}  -2 m \,\phi_{s} 
\right]_{r_{L}}^2
\end{equation} 
within the accuracy of integration, as is expected (\citet{shu1985a}).
Figure \ref{fig:torque} shows the scaled accumulated torque $T/T_{Lin}$ for a $m=2$ density wave with forcing parameter $\delta_{s} = 0.37$.

\begin{figure}[ht!]
\vspace{-0.2cm}
\centering
\subfloat[][]{\includegraphics[width = 0.55 \textwidth]{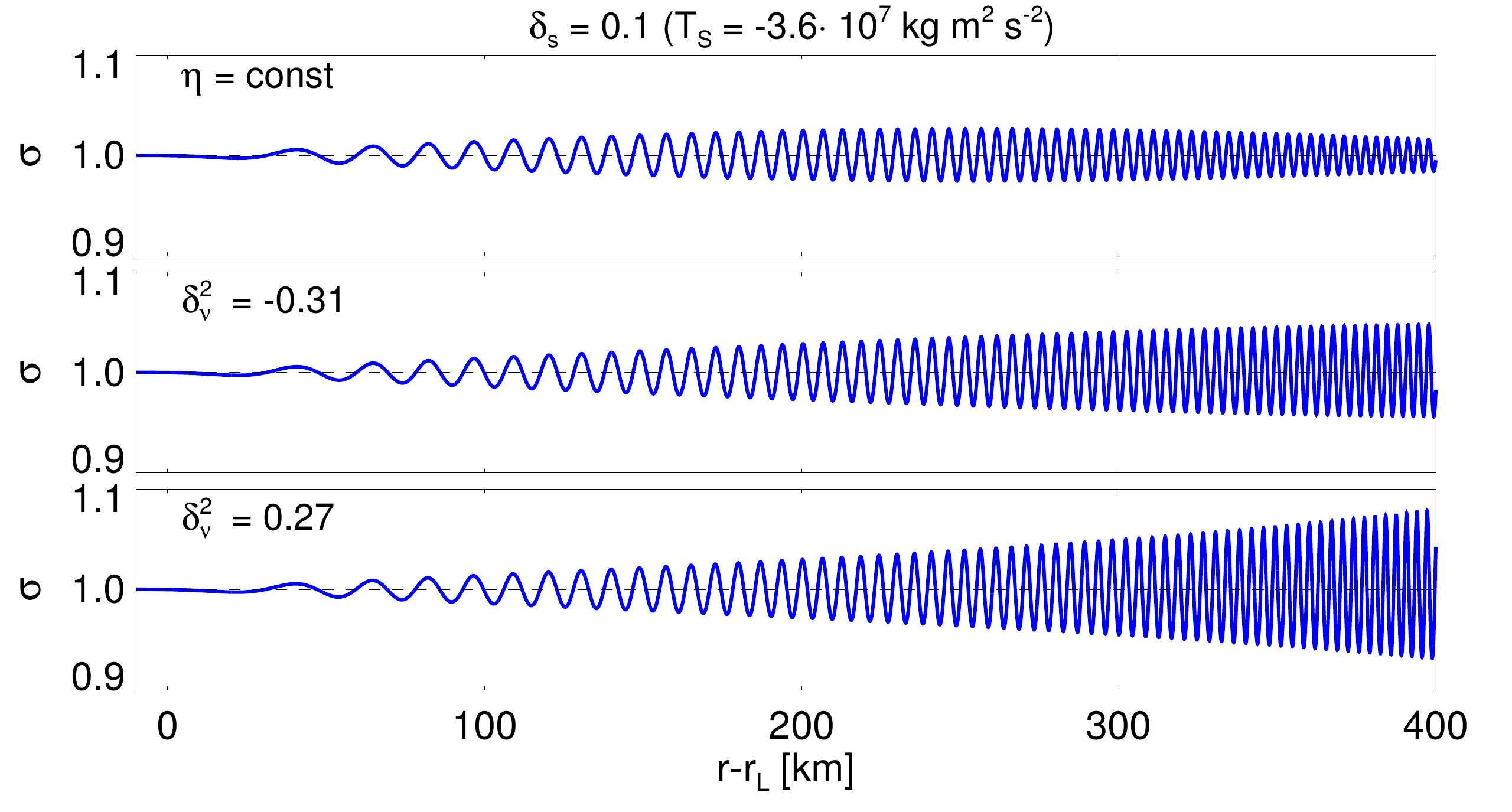}}\\
\vspace{-0.3cm}
\subfloat[][]{\includegraphics[width = 0.55 \textwidth]{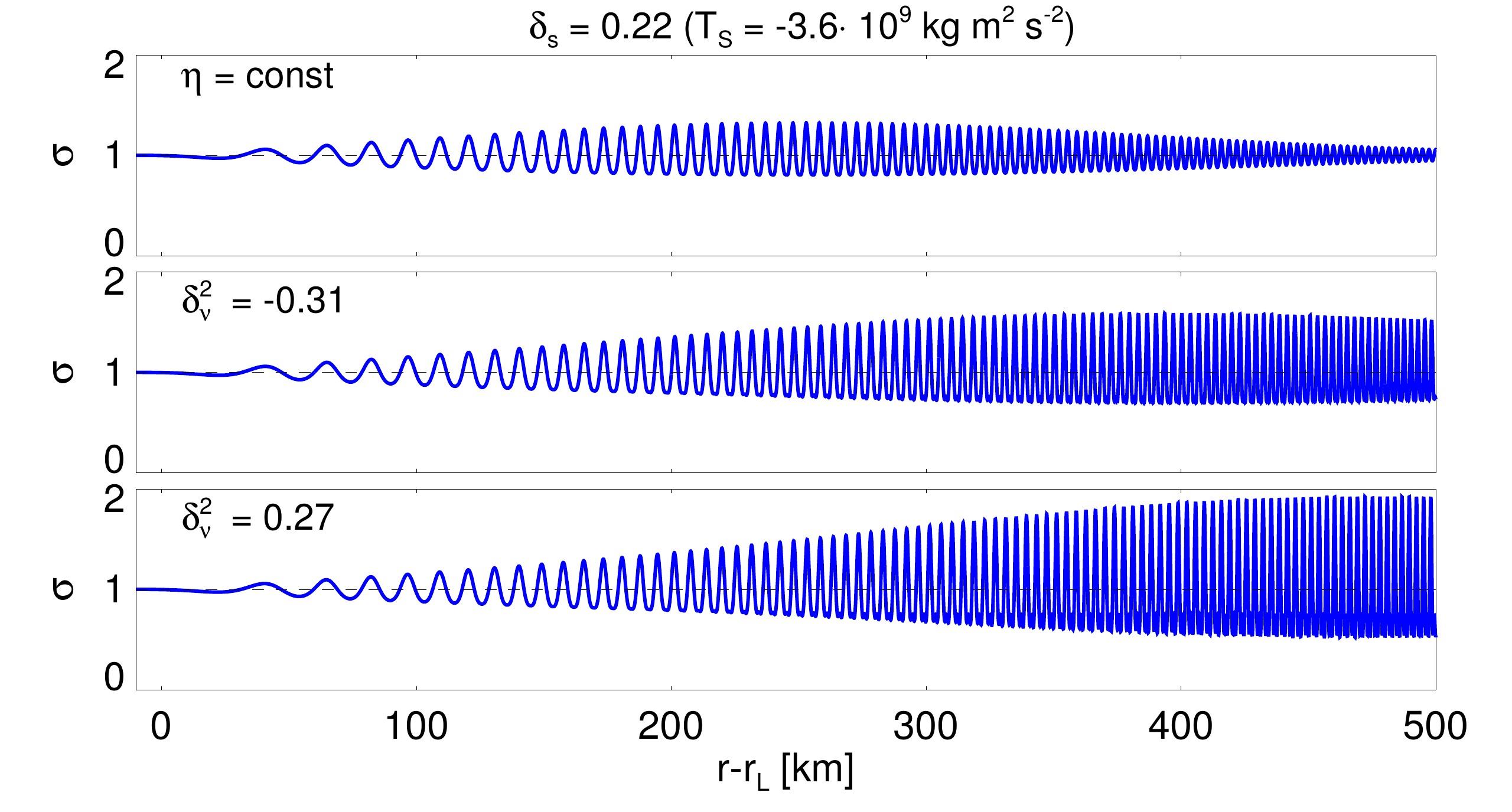}}\\
\vspace{-0.3cm}
\subfloat[][]{\includegraphics[width = 0.55 \textwidth]{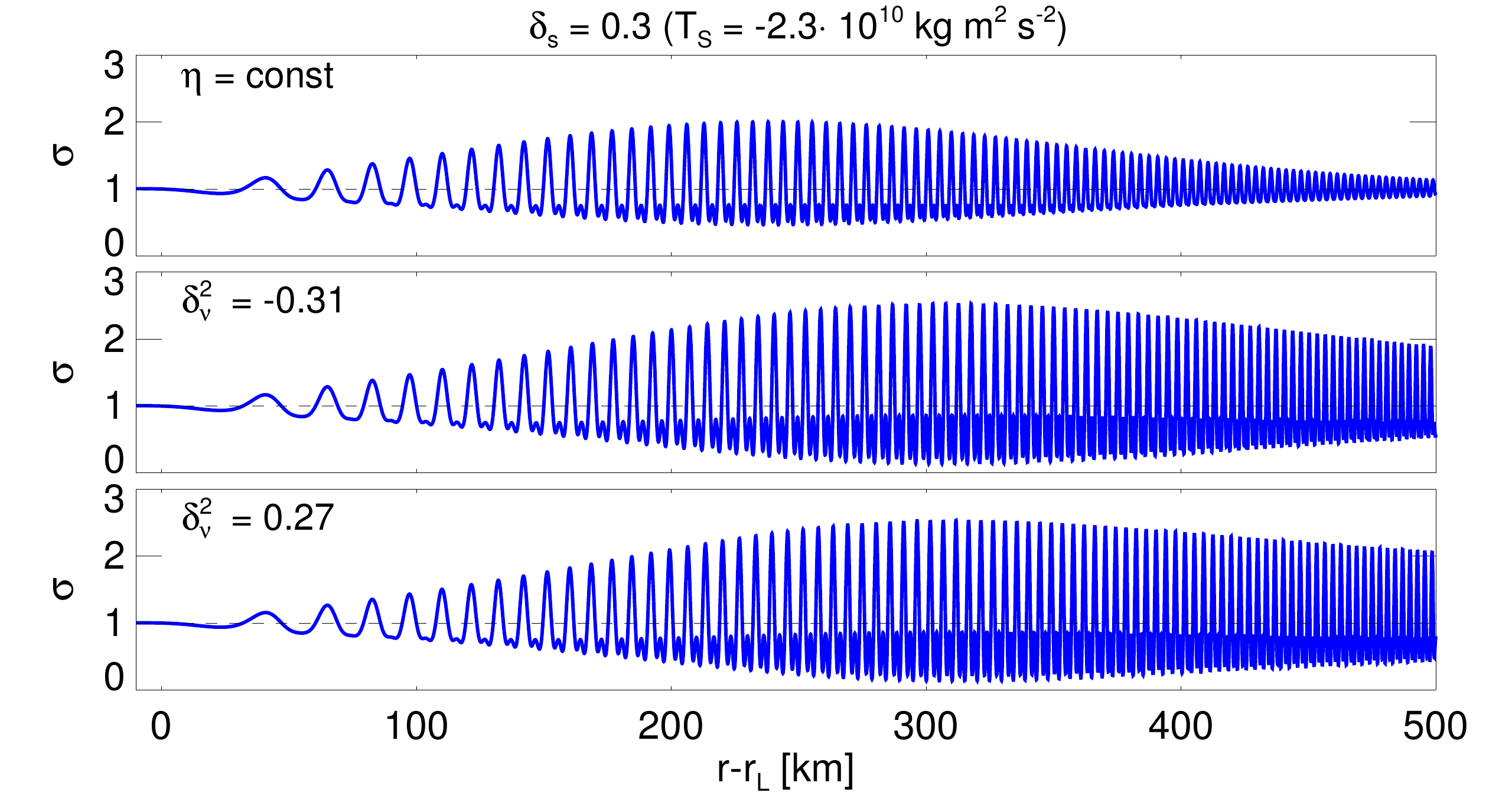}}
\vspace{-0.1cm}
\caption{Plots of $m=2$ density waves to illustrate the effects of increasing $\delta_{s}$ and $\delta_{\nu}$. 
     For each panel (a), (b) and (c), the profiles from top to bottom correspond to the regimes (1, 2, 4), 
     (1, 3, 4) and (1, 3, 4), respectively. For the first wave in each panel we used $\nu_{0}$ and $\gamma$ from the \emph{$\tau_{10}$}-parameters. 
     For the second and third waves we used the \emph{$\tau_{10}$}- and \emph{$\tau_{20}$}-parameters, respectively.
     Further, $r_{L}=96248\,\text{km}$ and $\sigma_{0}=600\, \text{kg}\,\text{m}^{-2}$ was used.}
\label{fig:fieldscompare}
\end{figure}

In analogy to the discussion at the end of Section \ref{sec:nlfree} we are interested in solutions of the amplitude equations (\ref{sub:fampeq1}), 
(\ref{sub:fampeq2}) which belong to the following parameter regimes
\begin{enumerate}
 \item Density waves, damped by constant viscosities ($\beta\to -1$).
 \item Linear, (marginally) stable density waves ($\beta  < \beta_{c}$), damped by density dependent viscosities.
 \item (Weakly) nonlinear density waves that are (marginally) stable ($\beta  < \beta_{c}$). Here the damping is caused both by linear and nonlinear terms.
  \item (Weakly) nonlinear\footnote{One should note that a wave of this category will eventually become nonlinear, regardless of the value of $\delta_{s}$. 
Therefore we do not distinguish linear from nonlinear waves in this case.} density waves that are (marginally) \emph{unstable} ($\beta  > \beta_{c}$). Here the 
damping is purely nonlinear and follows a power law behavior at large distances.
 \end{enumerate}
Case 1 is considered to illustrate the differences caused by the density dependence of viscosity on the wave damping and because such models have been applied 
to investigate density waves in Saturn's rings [\citet{esposito1983c};~\citet{tiscareno2007b};~\citet{colwell2009b}].
The wave profiles corresponding to case 1 are computed by solving the amplitude equations (\ref{sub:fampeq1}), (\ref{sub:fampeq2}) in the limit $\beta \to -1$ 
(cf. Section \ref{sec:lintheo}).
Therefore their damping is affected also by the nonlinear terms in Eqs. (\ref{sub:fampeq1}), (\ref{sub:fampeq2}) for sufficiently large amplitude.
The model parameters that separate the cases 2, 3 and 4 are essentially the strength of satellite forcing $\delta_{s}$  
and the distance of the ring state to the threshold for viscous overstability $\delta_{\nu}$. Values of $\delta_{s} \sim 1$ correspond to strongly nonlinear 
waves, whereas, if $\delta_{s} \ll1$, a stable wave remains in the linear regime.
Further, positive values $\delta_{\nu}^2>0$ imply (linear) viscous overstability while in the case $\delta_{\nu}^2<0$, the wave is viscously stable, which 
formally corresponds to an \emph{imaginary value of} $\delta_{\nu}$.
To illustrate the effects of different values of $\delta_{s}$ and $\delta_{\nu}$, Figures \ref{fig:fieldscompare} a, b and c show wave profiles for the cases 1, 
2, 3 and 4.  
In these plots the waves damped by constant viscosity have substantially shorter damping lengths than those damped by density dependent viscosity.
However, for other choices of model parameters the upper two waves in Figures \ref{fig:fieldscompare} a, b, c, respectively may look fairly similar.

\section{Comparison with the Model of BGT86}\label{sec:bgtcomp}
In this section we compare the derived weakly nonlinear model (WNL hereafter) with the nonlinear streamline model of \citet{bgt1986} (BGT86 hereafter).
To this end we compute density wave profiles by applying the method outlined in Section IVa in BGT86 where we use the viscosity prescription 
(\ref{eq:shearvis}).
We perform these calculations to leading order of the parameter $x$ [defined by (\ref{eq:xx})], similar to our calculations in Sections \ref{sec:nlfree} and 
\ref{sec:nlforced}.
In Appendix \ref{sec:ptensor} we derive the pressure tensor components and the viscous coefficients used in the streamline model.  

\begin{figure}[ht!]
     \centerline{\includegraphics[width=0.75 \textwidth]{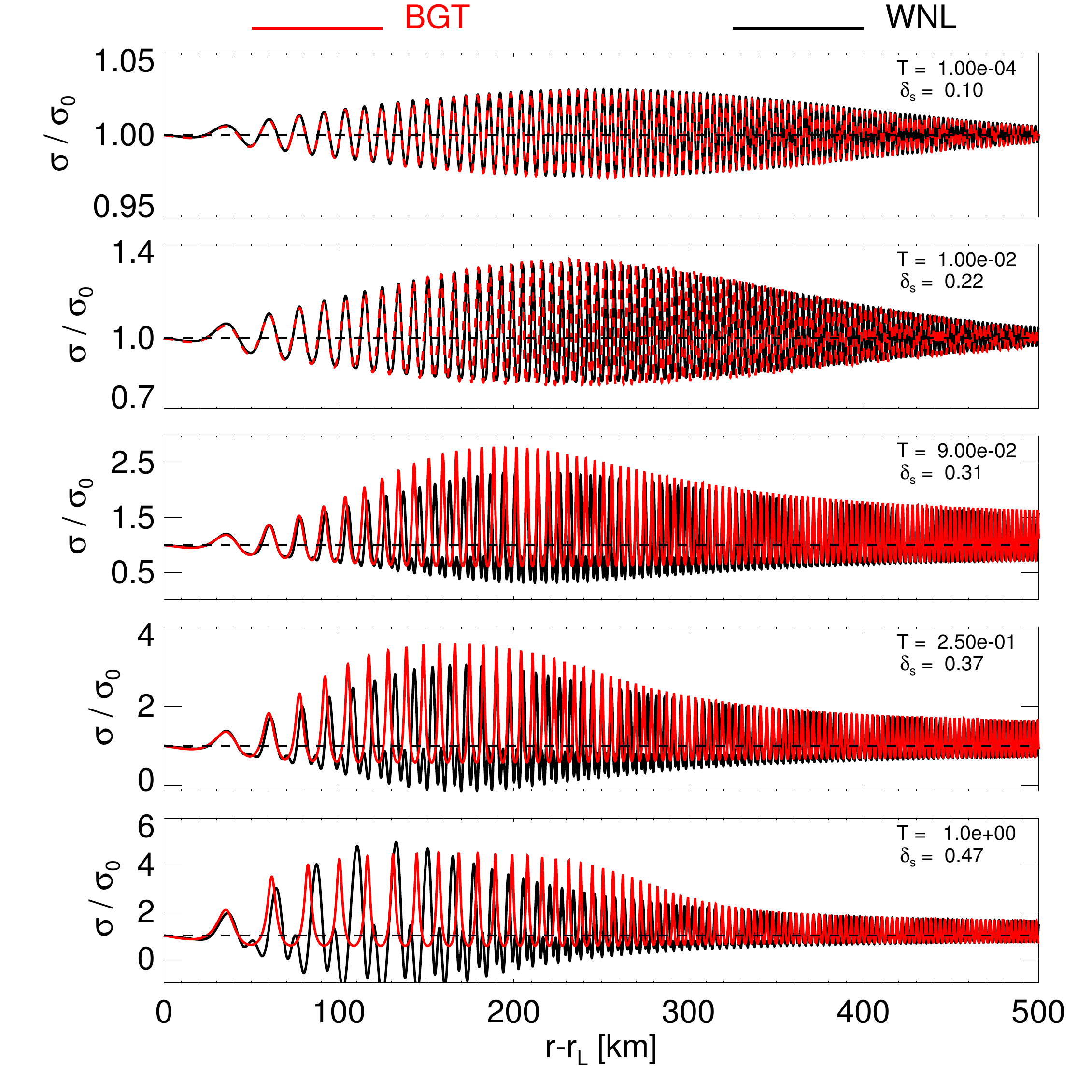}}
     \caption{Density wave profiles following from the WNL model and the BGT model respectively. Forcing strengths of the waves, expressed through 
$\delta_{s}$ and the scaled torque $T$, increase gradually from top to bottom.
     The BGT profiles have been divided by the corresponding background surface densities (Fig. \ref{fig:sigmabg}).
     In the two lowermost cases the surface densities of the WNL model waves become negative where the amplitudes are largest. This is a consequence of the 
limitation of the WNL model to 
     second order harmonics of the primary wave solution (Section \ref{sec:ampeq}). For a quantitative description of the wave profiles of these strongly forced 
waves more
     higher harmonics must be included. Nevertheless, the amplitude profile derived by the model remains valid and is unaffected by the restriction on second 
order harmonics.}
     	\label{fig:januscompare}
\end{figure}

Figure \ref{fig:januscompare} shows wave profiles derived from both models. From top to bottom panel the forcing strength $\delta_{s}$ [Eq. (\ref{eq:deltafor})] 
gradually increases from the linear to the strongly nonlinear regime. 
The given torques $T$ are scaled with the linear inviscid torque value of the Janus 2:1 resonance [$T_{S}= -3.61\cdot 10^{11} \, \text{kg}\,\text{m}^2 
\,\text{s}^{-2}$, obtained from (\ref{eq:lintor})], which corresponds to a strongly nonlinear wave.
The satellite torques serve as input parameters for the BGT model, whereas the wave forcing in the WNL model is computed from Eq. (\ref{eq:fampeq}).
For all plots we use $r_{L}=96248\,\text{km}$, $\sigma_{0}=600\, \text{kg}\,\text{m}^{-2}$ and $m=2$.
For the three lowermost plots we use the $\tau_{15}$-parameters, but with a higher viscosity $\nu_{0}=0.0025 \, \text{m}^2 \, \text{s}^{-1}$. 
This value is slightly higher than what is predicted by relation (\ref{eq:daisaka}) for ice particles at this saturnocentric distance, to ensure convergence of 
the solution procedure for equations (\ref{sub:fampeq1}), (\ref{sub:fampeq2}) for the lowermost case with $T=1$. For the two uppermost plots we use the 
$\tau_{10}$-parameters with $\nu_{0}=0.0025 \, \text{m}^2 \, \text{s}^{-1}$. 

Throughout the linear and weakly nonlinear regime we observe good agreement between the models. 
However, with increasing forcing strength, the deviations become larger.
This becomes most clear for the case $T=1$. Here we observe significant deviations in the region of maximal wave amplitudes. For the cases $T=0.25$ and $T=1$ 
one encounters negative values for the surface mass densities in the WNL profiles.
This indicates that the weakly nonlinear description breaks down in the regions of maximal wave amplitudes for these cases (but higher order corrections might 
remedy this shortcoming, see the discussion at the end of Section \ref{sec:ampeq}).
One should note the different plot ranges used in the different panels.
In contrast, the saturation amplitudes, i.e. the near constant amplitudes at far distances from resonance in the cases with viscous overstability, agree very 
well, with a deviation of about $3\%$ (not shown) for the strongest waves with $T=1$.
This is remarkable, considering that the approaches behind the two models are entirely different. 

\begin{figure*}[ht!] 
\vspace{-1cm}
     \centerline{\includegraphics[width=0.9 \textwidth]{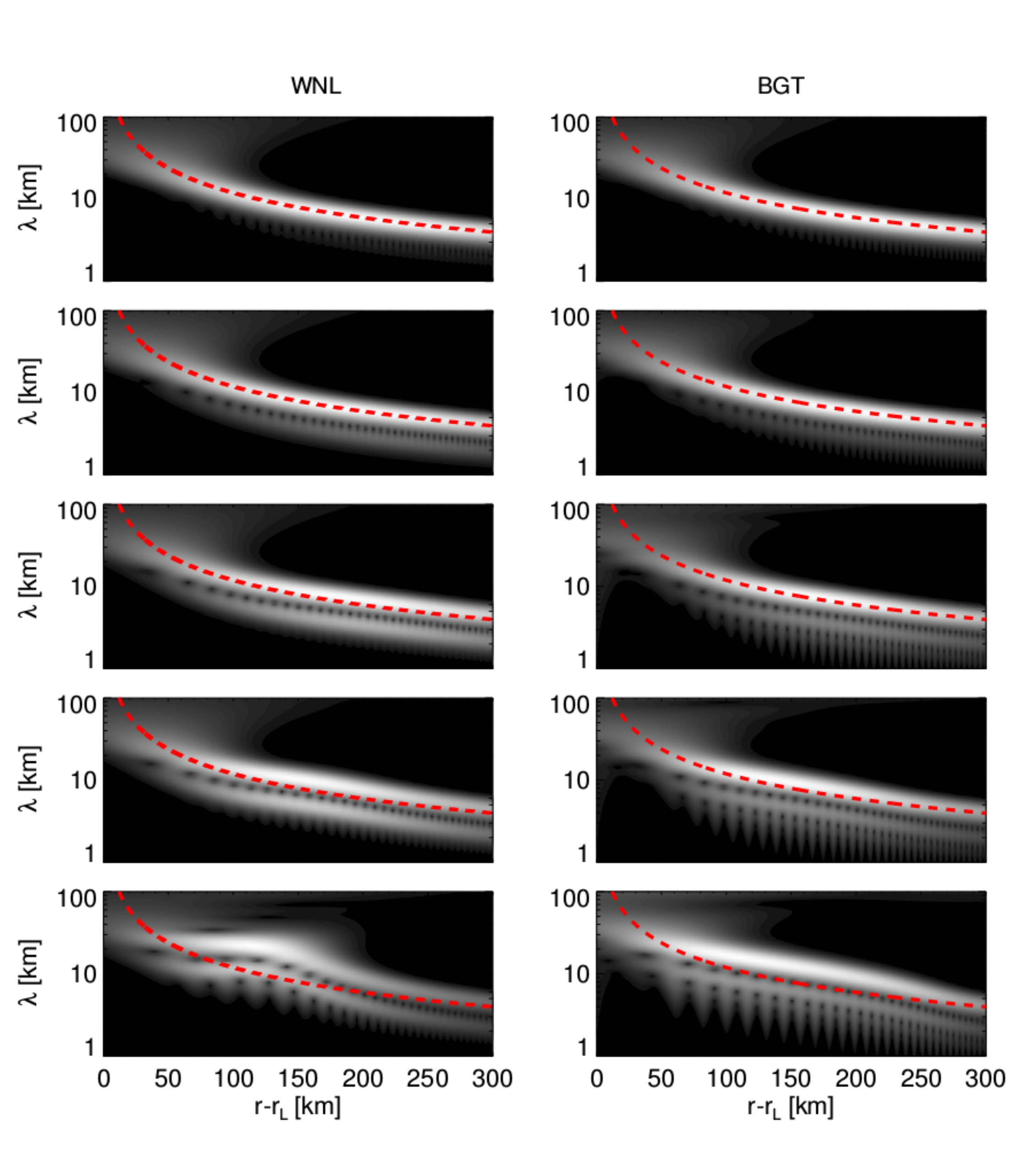}}
     \caption{Morlet wavelet spectrograms of the model waves presented in Fig. \ref{fig:januscompare}. The dashed line represents the linear 
dispersion relation $k=\frac{\mathcal{D}x}{2\pi G \sigma_{0}}$ for reference. 
     As the satellite torque increases from top to bottom panels, departures from the linear dispersion relation increase. As already seen in Figure 
\ref{fig:disprel}, nonlinearity tends to increase the wavelength of density waves.
     Similar to the wave amplitudes, we observe that differences between the models become larger on a quantitative level, as one enters the strongly nonlinear 
regime.}
     	\label{fig:januscomparedisp}
\end{figure*}

\begin{figure*}[ht!]
     \centerline{\includegraphics[width=0.45 \textwidth]{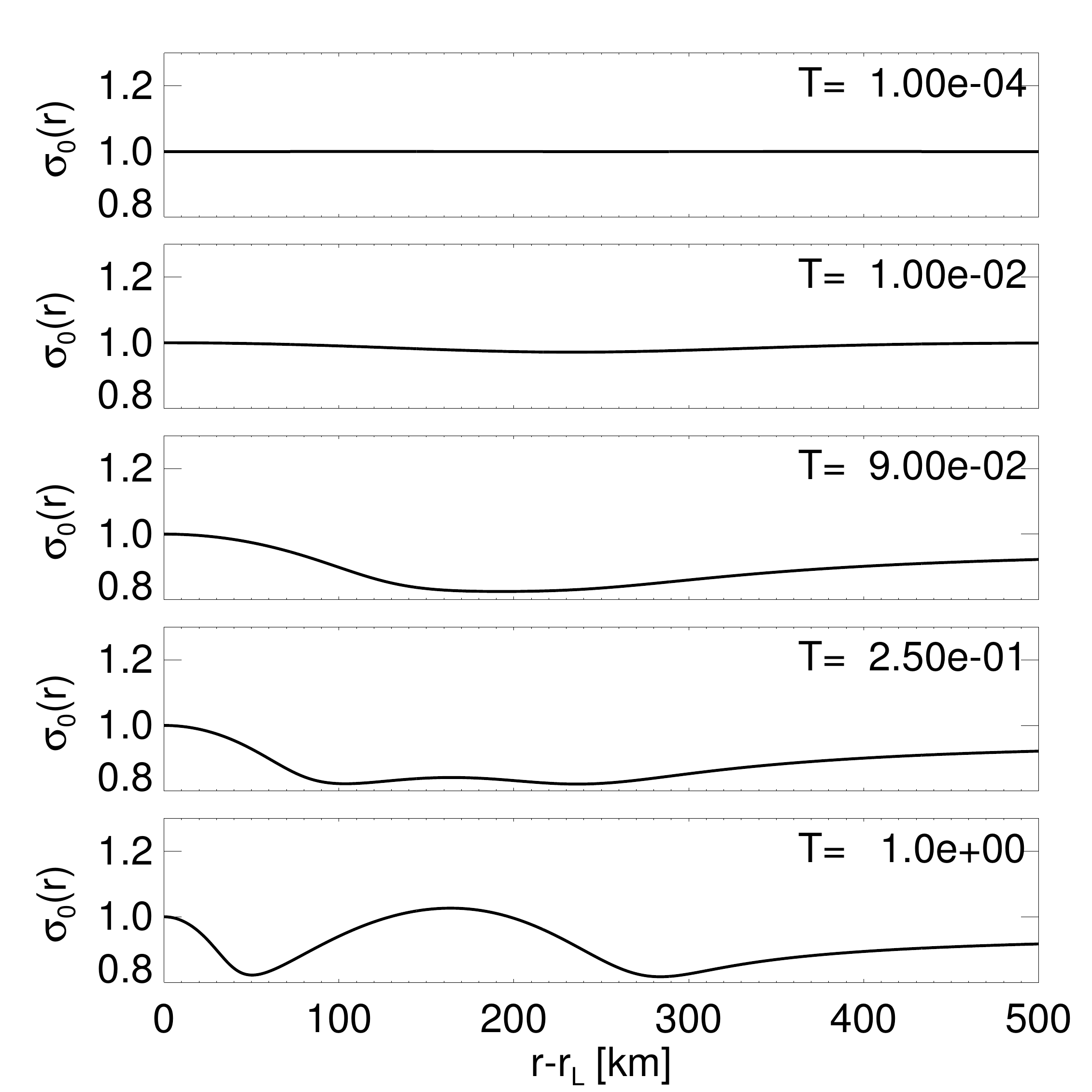}}
     \caption{Background surface density profiles for the waves of the BGT model displayed in Figure \ref{fig:januscompare}.
			These curves are scaled with the constant $\sigma_{0}=600\, \text{kg}\,\text{m}^{-2}$.}
     	\label{fig:sigmabg}
\end{figure*}

Apart from the differences in the amplitude profiles, the waves also exhibit different wavenumber dispersions. In Figure \ref{fig:januscomparedisp}  we present 
Morlet wavelet spectrograms (\citet{torrence1998}) of the waves displayed in Fig. \ref{fig:januscompare}.
In the linear and weakly nonlinear regime, the wave numbers from both models closely follow the linear dispersion relation. In the nonlinear regime, both models 
predict longer wavelengths in regions of large wave amplitudes (cf. Figure \ref{fig:disprel} in Section \ref{sec:ampeq}).
To some extent, differences between the dispersion relations and density profiles of the two models arise because in the BGT model the background surface 
density changes with radial position, due to
the forced conservation of angular momentum luminosity. This effect is not included in the WNL model.
Therefore the BGT density wave profiles in Figure \ref{fig:januscompare} are scaled with the corresponding background surface densities $\sigma_{0}(r)$,
whereas the WNL model waves are scaled with the constant $\sigma_{0}=600\, \text{kg}\,\text{m}^{-2}$.
The background densities for the BGT model waves are shown in Figure \ref{fig:sigmabg}. 

\begin{figure*}[hb!] 
     \centerline{\includegraphics[width=0.45 \textwidth]{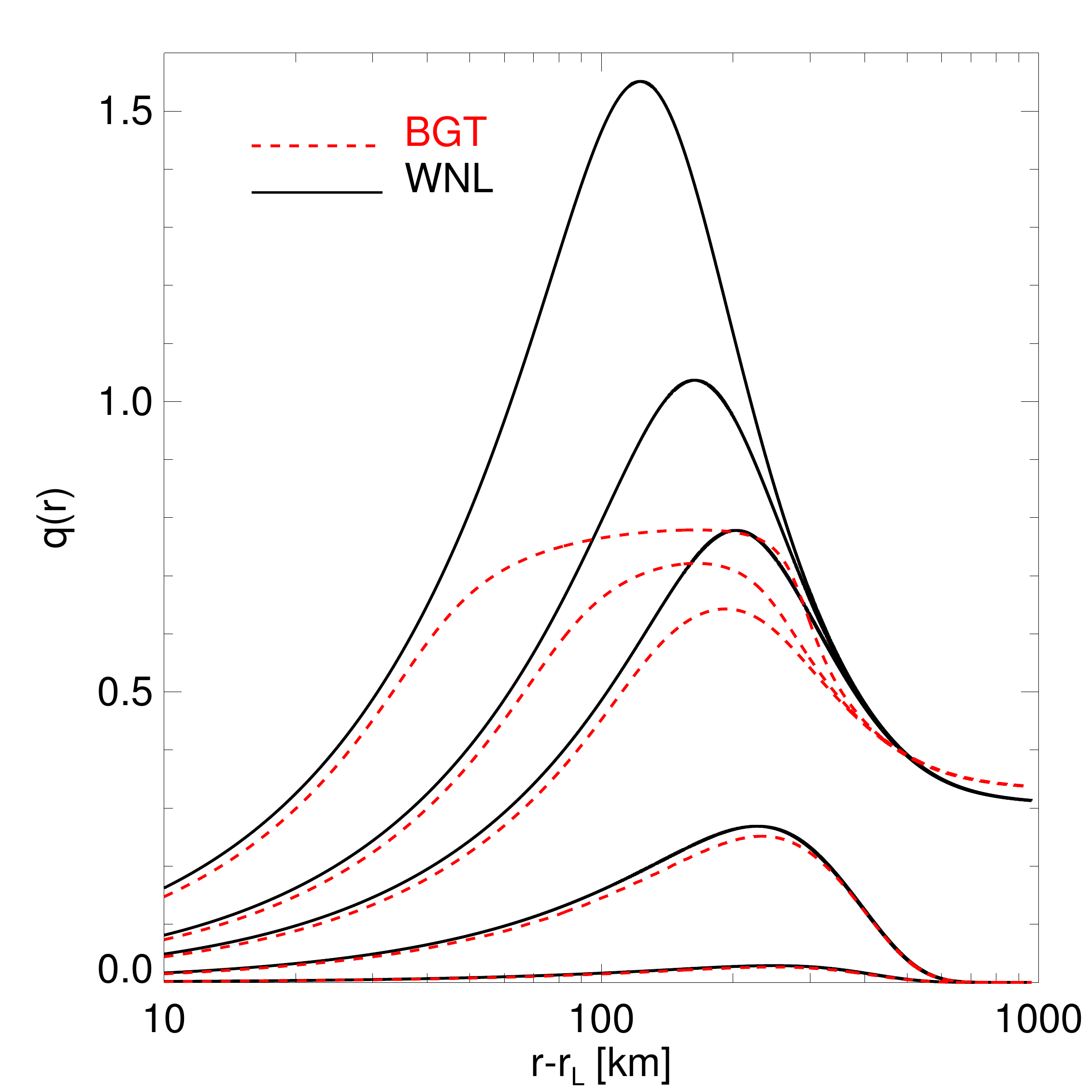}}
     \caption{Nonlinearity parameters for the waves displayed in Figure \ref{fig:januscompare}. Eq. (\ref{eq:qpar}) was used for the waves of the WNL 
model. 
     Curves with higher values of $q$ correspond to higher values of $\delta_{s}$ and $T$. Note that in regions where $q>1$ the WNL model breaks down and 
relation (\ref{eq:qpar})
     is no longer a valid approximation for $q$ (see Appendix \ref{sec:ptensor}).}
     	\label{fig:qpar}
\end{figure*}

Further, Figure \ref{fig:qpar} displays the nonlinearity parameter $q$ as a function of radial distance from resonance for the waves.	
We use (\ref{eq:qpar}) to obtain $q$ for the WNL model waves. 
For the cases of overstable waves, $q$ saturates to a finite value. 
This saturation value is determined by the condition that the radial derivative of the angular momentum
luminosity $\mathrm{d} L / \mathrm{d}x$, carried by the density wave, approaches zero at large distances. 
In the WNL model, this results in the condition that the two viscous terms in (\ref{eq:lflux}) balance each other, which is fulfilled 
if $|\mathcal{A}|$ approaches the limiting function (\ref{sub:powerlawdamp}). This occurs at large distances $x$ where
the influence of the satellite torque is negligible.
In the BGT model, the wave damping is described through [cf. (27) in BGT86]
 \begin{equation}
 \frac{\mathrm{d}L}{\mathrm{d}x} = -2 \pi \, m q \, \sigma_{0}(x)  \mathcal{T}_{1}(q)
\end{equation} 
with the viscous coefficient $\mathcal{T}_{1}(q)$ [Eq. (17) in BGT86 and (\ref{eq:viscous_coefficient_1}) in Appendix \ref{sec:ptensor}].
 In the case of viscous overstability ($\beta>\beta_{c}$) the quantity $\mathcal{T}_{1}$ is positive for $0< q < q_{c}$ with some
finite value $q_{c}$. This generates a linear instability of the density wave in a similar way as the linear terms $\propto |\mathcal{A}|$ 
in the amplitude equations (\ref{eq:landauampfree}) and (\ref{eq:fampeq}) if the condition $\beta>\beta_{c}$ is met. The critical value $q_{c}$ is approached as 
$x$ becomes larger. For the $\tau_{15}$-parameters one finds $q_{c}=0.330$ in the BGT model. 
This behavior can be compared with the fact that the saturation amplitude (\ref{sub:powerlawdamp}) of the WNL model
for which $\mathrm{d} L / \mathrm{d}x \to 0$ corresponds to a $q$ value of $4 \sqrt{\delta_{\nu}^2 \, g_{r}/l_{r}}$, yielding $q_{c}=0.30$ for the 
$\tau_{15}$-parameters. The good agreement of these values for $q_{c}$ is reflected in the agreement of the saturation amplitudes of the overstable density 
waves in Fig. \ref{fig:januscompare}.

One should note that the two most nonlinear cases in this comparison are strictly speaking outside the regime of applicability of the weakly nonlinear model,
the surface density even becoming negative in some regions.  The two strongest waves possess values for $q$ greater than unity within some regions. This shows 
that in these cases, a second order description of the density perturbation is not sufficient to quantitatively
describe the wave profile.
Nevertheless, we see that for all cases the wave envelope, which is the central quantity of the weakly nonlinear model,
is in good or at least qualitative agreement with the BGT model, indicating that the amplitude equation remains qualitatively valid.

\newpage
\newpage
\newpage


\section{Summary and Discussion}\label{sec:summary}
In this paper we applied the multiple scale approach to derive a weakly nonlinear fluid model for the excitation and viscous damping of spiral density waves in 
a dense planetary ring.
The most important quantity obtained with this model is the evolution profile of the amplitude of a density wave depending on the distance from resonance 
location.
The model takes into account nonlinearities which are present in the governing fluid equations and which become important
if the density perturbations are of the same order as the background value. This is the case for many of the observed density waves in Saturn's main rings.

A linear instability of a density wave arises if the condition for viscous overstability is fulfilled (\citet{schmidt2016}).
We find that the damping of such overstable density waves occurs solely due to the nonlinear terms in the hydrodynamic balance equations. For large distance 
from the resonance we derive a power law damping for the scaled wave amplitude.
As a consequence, the surface mass density perturbation in the model saturates to a finite value. In a true particulate ring one expects that the wave 
eventually disappears far from resonance. 
In contrast, an exponential damping relation results from a linearized description with constant viscosity. In general, the resulting density wave 
damping lengths are strongly dependent on the distance from the threshold of viscous overstability. We believe that this dependence can in part explain
the wide variety of damping lengths observed among waves in Saturn's rings, such as the waves at the first order resonances with the co-orbitals 
(\citet{schmidt2016}). 

Our model predicts, in accordance with existing theories of nonlinear density waves, distinct features of strong waves in Saturn's rings.
Among these are the sharp peaks and flat troughs of the radial density profiles and the deviations from the linear dispersion relation in regions of strong wave 
amplitude.
Moreover, our calculations show that long range self-gravity contributions, not present in linear WKB-approximation, have the tendency to reduce damping 
lengths.  

The results from our new approach compare reasonably well with the traditional streamline approach to density wave theory. The largest deviations occur for 
strong forcing
and in the highly nonlinear regime. The streamline approach is superior at matching the total wave profile, while 
the newly derived amplitude equation in this paper is a comparably handy tool
to gain insight in the evolution of the wave amplitude with distance from resonance, and the different regimes of wave formation and the dependence on the 
parameters of the model.

A detailed quantitative reproduction of the observed strongly nonlinear waves in Saturn's rings can be achieved
with neither one of the models. Both models rely on a isothermal (vertically averaged) fluid approximation neglecting the effect of the wave on the local 
velocity dispersion
of ring material and variations of the ring thickness with the wave phase. Also, the magnitude of gravitational wakes (and their contribution to the viscosity) 
will be influenced in a more or less complicated manner by the presence of a density wave.
Further, we expect that a realistic description of self-gravity, which takes into account long range interactions more accurately, leads to further deviations 
from the model dispersion relation
discussed in this paper. 

ring regions
simulation.

Overstability in the inner B ring might explain the remarkable length ($>500 \,\text{km}$) of the Janus 2:1 wave train (\citet{colwell2009}).
On the other hand, visible wave signatures extend to less than about $200 \,\text{km}$ for 
two waves propagating in the overstable region of the A ring (Atlas 7:6 and Pan 10:9, \citet{hedman2014a} Figure 5).
This appears surprising in view of the results from our weakly nonlinear model and from the BGT model.
However, in this regard it should be noted that although the models describe the behavior of a density wave if the condition for viscous overstability is 
fulfilled, they do
not take into account the presence of overstable oscillations in the wave region. A study of the interplay of these wave phenomena requires a numerical 
simulation.

\section*{Acknowledgments}

We acknowledge support from the Academy of Finland and the University of Oulu Graduate School.
We thank Frank Spahn for valuable discussions and an anonymous reviewer for a constructive report that helped us to greatly improve the 
paper.




\appendix

\section{The Nonlinear Terms $N_{2}$ and $N_{3}$}\label{sec:nlterms}

The nonlinear Terms $N_{2}$ and $N_{3}$ appearing in Eqs. (\ref{eq:eqnexp}) are given by

 \begin{equation}
  \mathbf{N_{2}}(\mathbf{\Psi_{1}},\mathbf{\Psi_{1}})   =  \begin{pmatrix} \mathbf{N_{21}} \\[0.1cm]
\mathbf{N_{22}}  \\[0.1cm] 
\mathbf{N_{23}}  \end{pmatrix}
\end{equation}
 with 
 \begin{align}
  \begin{split}
 \mathbf{N_{21}}& = \frac{\epsilon k}{\mathcal{D}} \partial_{x} \left( u_{1} \phi_{1} \right),\\
 \mathbf{N_{22}}& =-\epsilon u_{1} \partial_{x} u_{1} +\frac{c^2 \epsilon k^2}{\mathcal{D}^2}\phi_{1} \partial_{x} \phi_{1}\\
 & \quad  - \frac{\epsilon^2 \alpha k \nu_{0}}{ \mathcal{D}} \left(\left[\beta_{c}+1\right] \partial_{x} \phi_{1} \partial_{x} u_{1} 
 + \beta_{c} \phi_{1} \partial_{x}^2 u_{1}\right),\\
 \mathbf{N_{23}}& =-\epsilon u_{1} \partial_{x} v_{1}\\
 & \quad -\frac{\epsilon k \nu_{0}}{2 \mathcal{D}^2} \left(\partial_{x}
 \phi_{1} \left(3\left[\beta_{c}^2 -1 \right]k \Omega \phi_{1}
 +\mathcal{D}\left(-3\beta_{1}\Omega + 2\left[\beta_{c}+1\right]\epsilon \partial_{x} v_{1}\right)\right)+2\beta_{c}\mathcal{D}\epsilon 
 \phi_{1} \partial_{x}^2v_{1}\right)
 \end{split}
 \end{align}

and
\begin{equation}
  \mathbf{N_{3}}(\mathbf{\Psi_{1}},\mathbf{\Psi_{2}})   =  \begin{pmatrix} \mathbf{N_{31}} \\[0.1cm]
\mathbf{N_{32}}  \\[0.1cm] 
\mathbf{N_{33}}  \end{pmatrix}
\end{equation}
with
\begin{align}
\begin{split}
  \mathbf{N_{31}} &=\frac{\epsilon}{\mathcal{D}} \left(-\mathcal{D} \partial_{\xi}u_{1} 
-i\Omega\partial_{\theta}\partial_{\xi}\phi_{1}-i\partial_{t}\partial_{\xi}\phi_{1} \right)
 +  k \left(\partial_{x}\left[u_{2} \phi_{1}\right] +2  \partial_{x}\left[u_{1}\phi_{2}\right] \right) ,\\
  \mathbf{N_{32}}   & =-\epsilon \left( \partial_ {\xi} \phi_{1} +\partial_{x}\left[u_{2}u_{1}\right] \right)\\
  \quad & +\frac{c^2 \epsilon}{\mathcal{D}^3} \left(\mathcal{D}^2 k \partial_{\xi}\phi_{1} + k^3 \phi_{1}^2 \partial_{x}\phi_{1} + 2 \mathcal{D}k^2 \phi_{1} 
\partial_{x}\phi_{2} 
  + \mathcal{D}\left(2 k^2 \phi_{2}\partial_{x}\phi_{1} -i \mathcal{D}\epsilon\partial_{x}\partial_{\xi}\phi_{1}\right)\right)\\
    \quad & + \nu_{0}\left(-\frac{\epsilon^2 \alpha k}{\mathcal{D}^2}\left(2\left[\beta_{c}+1\right]\mathcal{D}\partial_{x}\phi_{2}\partial_{x}u_{1} 
+\partial_{x}\phi_{1}\left(
  \left(\beta_{1}\mathcal{D}+k \left[1-\beta_{c}^2 \right]\phi_{1}\right)\partial_{x}u_{1} + 
\left[1+\beta_{c}\right]\mathcal{D}\partial_{x}u_{2}\right)\right)\right.\\
  \quad & \left. +\epsilon^2 \alpha\left(2 \partial_{x}\partial_{\xi}u_{1} + 
\frac{k}{2\mathcal{D}^2}\left(-2\beta_{1}\mathcal{D}\phi_{1}+\left[\beta_{c}-1\right]\beta_{c}
  k\phi_{1}^2 - 4 \beta_{c} \mathcal{D} \phi_{2}\right)\partial_{x}^2 u_{1} -\frac{\beta_{c} k}{\mathcal{D}}\phi_{1}\partial_{x}^2 u_{2}\right)\right),\\
 \mathbf{N_{33}}  & = -\epsilon\left(u_{2}\partial_{x}v_{1} + u_{1} \partial_{x} v_{2}\right)\\
  \quad & + \frac{\nu_{0} \epsilon}{4 \mathcal{D}^3}\left(6\left[1+\beta_{c}\right]\mathcal{D}^2 k\Omega \partial_{\xi}\phi_{1}+k\partial_{x}\phi_{1}
  \left(3\left[\beta_{c}-2\right]\left[\beta_{c}^2 -1 \right] k^2 \Omega \phi_{1}^2\right.\right. \\
  \quad & + 4 \mathcal{D} k \phi_{1}\left(-3 \beta_{1}\beta_{c}\Omega
  +\left[\beta_{c}^2 -1\right] \epsilon \partial_{x}v_{1}\right) + 2 \mathcal{D}\left(-6 \left[\beta_{c}^2 -1\right] k \Omega \phi_{2}\right. \\
  \quad & \left.\left. + \mathcal{D} \left(3 \mathrm{sgn}\left(\delta_{\nu}^2 \right) \beta_{2} \Omega   -2\beta_{1} \epsilon \partial_{x}v_{1} - 
2\left[\beta_{c}+1\right]\epsilon \partial_{x} v_{2}\right)\right)\right)\\
 \quad &  - 2 \mathcal{D}\left(2 k \partial_{x}\phi_{2} \left(
  -3 \beta_{1} \mathcal{D} \Omega + 3 \left[\beta_{c}^2 -1\right]k \Omega \phi_{1} + 2 \left[1+\beta_{c}\right] \mathcal{D}\epsilon 
\partial_{x}v_{1}\right)\right.\\
  \quad & +\epsilon \left(3 i \left[\beta_{c}+1\right]\mathcal{D}\Omega \partial_{x}\partial_{\xi}\phi_{1} -4 \mathcal{D}^2 \partial_{x}\partial_{\xi}v_{1} + 
  k\left(\phi_{1} \left(2 \beta_{1} \mathcal{D} - \left[\beta_{c}-1\right]\beta_{c} k \phi_{1} \right) + 4 \beta_{c} \mathcal{D} \phi_{2} \right)\partial_{x}^2 
v_{1}\right.\\
  \quad & \left.\left.\left. +2\beta_{c}\mathcal{D} k \phi_{1} \partial_{x}^2 v_{2}\right)\right)\right).
    \end{split}  
\end{align}
In the above expressions we defined the constant $\alpha=4/3+\gamma$.
Note that the solvability condition (\ref{eq:beta1}) leads to a simplification of above terms.

\newpage

\section{Solution of the Poisson Equation}\label{sec:poisson}

To conduct the multiple scale expansion of the nonlinear fluid equations (\ref{eq:nleq}) one needs to find the relationship between disk potential $\phi_{i}$ 
and disk surface density $\sigma_{i}$
for the multiple scale orders $i=1,2,3$.
Poisson's equation for a thin disk reads (in unscaled form)
 \begin{equation}\label{eq:poisson1}
\frac{1}{r}\, \frac{\partial}{\partial r}\left(r\, \frac{\partial \phi}{\partial r}\right)+\frac{1}{r^2}\frac{\partial^{2}\phi }{\partial \theta^{2}}+ 
\frac{\partial^{2} \phi}{\partial z^{2}}     = 4 \pi G \sigma \delta (z).
\end{equation}
with the Dirac delta function $\delta\left(z\right)$. 
If the surface density has the form
 \begin{equation}\label{eq:sigmaperturba}
\sigma(r,\theta,t) = A\left(r\right) \cdot \exp\Bigg\{i \int_{}^{r} k \left(\hat{r}\right)\, \mathrm{d}\hat{r} \Bigg\}  \cdot \exp \left\{i\left(m\theta-\omega 
t\right) \right\},
\end{equation} 
with rapidly varying phase, such that $|k A| \gg \partial_{r}A$, we can neglect any curvature terms in (\ref{eq:poisson1}) [which correspond to corrections on 
the order of the small parameter $(kr)^{-1}$] and we are left with
 \begin{equation}\label{eq:poisson2}
\frac{\partial^{2} \phi}{\partial r^{2}} + \frac{\partial^{2} \phi}{\partial z^{2}}     = 4 \pi G \sigma \delta (z).
\end{equation} 
The solution of (\ref{eq:poisson2}) fulfilling the correct boundary conditions ($\phi\rightarrow 0$ for $|z|\rightarrow \infty$) is given by (\citet{Shu1970})
 \begin{equation}\label{eq:poissonsol}
 \sigma \left(r\right) =\frac{i\,\mathrm{s}}{2 \pi G} \, \frac{\partial \phi}{\partial r},
\end{equation} 
in the plane $z=0$ where $s=\text{sgn}\left(k\right)$ is the sign of the wavenumber $k$ in (\ref{eq:sigmaperturba}).
We can now introduce the expansions (\ref{sub:msexp}) and (\ref{eq:slowlength}) in (\ref{eq:poissonsol}) and collect different orders in $|\delta_{\nu}|$.
From this directly follows (in dimensionless form)
\begin{subequations}
  \begin{align}
  \sigma_{1} (x) &  = \frac{i \mathrm{s}\epsilon}{\mathcal{D} } \, \frac{\partial \phi_{1}}{\partial x}\label{sub:poisson1},\\[0.1cm]
\sigma_{2} (x) & =\frac{i \mathrm{s}\epsilon}{\mathcal{D} } \, \frac{\partial \phi_{2}}{\partial x}\label{sub:poisson2},\\[0.1cm]
\sigma_{3} (x)& = \frac{i \mathrm{s}\epsilon}{\mathcal{D} } \frac{\partial \phi_{3}}{\partial x} + \frac{i  \mathrm{s} \epsilon }{\mathcal{D} }  \frac{\partial 
\phi_{1}}{\partial \xi},\label{sub:poisson3}
 \end{align} 
\end{subequations}
with $\mathcal{D}=3\,(m-1)$. The second term in (\ref{sub:poisson3}) accounts for the effect of the variation of the potential (with the wave amplitude) on the 
length scale $\xi$.\\

For the solution procedures of the first and second order equations in (\ref{eq:eqnexp}) which are presented in Sections \ref{sec:linstab} and 
\ref{sec:secondorder}, 
respectively, one needs to evaluate the terms $\mathrm{d}\sigma / \mathrm{d}\phi$ as these appear in the 
linear operator (\ref{eq:linop}) and its adjoint (\ref{eq:linopad}).
From expression (\ref{eq:psi1}) for the first order vector of state directly follows $\partial \phi_{1}/\partial x=\frac{i k}{\epsilon}\phi_{1}$, so that 
with (\ref{sub:poisson1}) we find $\mathrm{d}\sigma/\mathrm{d}\phi=-\frac{k}{\mathcal{D}}$ for the first order equations. 
To solve the second order equations in (\ref{eq:eqnexp}) we make the (natural) assumption that the second order self gravity potential $\phi_{2}$ is purely 
oscillatory and that it consists only of
the second harmonic of the first order potential $\phi_{1}$, i.e.
\begin{equation*}
 \phi_{2}\sim \mathcal{A}(\xi)^{2} \exp \left\{2 i\,\left[ \int  \frac{k}{\epsilon} \, \mathrm{d}x-\omega t +m\theta\right] \right\} .
\end{equation*}
With this assumption one finds $\partial \phi_{2}/\partial x=\frac{2 i k}{\epsilon}\phi_{2}$ such hat $\mathrm{d}\sigma/\mathrm{d}\phi=-\frac{2 k}{\mathcal{D}}$ 
is to be used
for the second order equations $\mathit{\hat{L}} \mathbf{\Psi_{2}}  = \mathbf{N_{2}}(\mathbf{\Psi_{1}},\mathbf{\Psi_{1}})$.
One notes that the so derived second order solution [Eqs. (\ref{eq:psi2}) and (\ref{eq:psi2sol})], is consistent with this assumption.
We do not attempt to find the corresponding expressions for the third order equations since these are not required to solve (\ref{eq:n3}) from which we obtain
the amplitude equation (\ref{eq:landauampfreescaled}).

Since Poisson's equation is linear, it applies to different Fourier-modes $\exp\left\{j \cdot  \int i \frac{k}{\epsilon}\, \mathrm{d}x\right\}$ separately with 
$j=\pm 1,\pm2,\pm3$.
Hence, all relations that result from solving Poisson's equation apply to isolated modes.
In this paper we use $s=1$, since we restrict our analysis to trailing density waves with $k>0$.

\section{Analytical Solution of the Amplitude Equation}\label{sec:ampsol}

Before we attempt to find the solution of the initial value problem 
 \begin{equation}\label{eq:landauamp2}
\frac{\mathrm{d}|\mathcal{A}|}{\mathrm{d}x} = \delta_{\nu}^{2} g_{r}\left(x\right) \, |\mathcal{A}| - l_{r}\left(x\right) \,|\mathcal{A}|^{3}
\end{equation} 
with $|\mathcal{A}|(x=0)=|\mathcal{A}_{0}|$ (derived in Section \ref{sec:nlfree}),
we can assess its asymptotic behavior by computing the amplitudes corresponding to fixed points. 
These are obtained by solving
 \begin{equation}
0 = \delta_{\nu}^{2} g_{r}\left(x\right) \, |\mathcal{A}| - l_{r}\left(x\right) \,|\mathcal{A}|^{3}.
\end{equation} 
As in Section \ref{sec:ampeq} we write $g_{r}\left(x\right)\equiv \hat{g}_{r} x^2$ and $l_{r}\left(x\right) \equiv \hat{l}_{r} x^4$ and note that 
$\hat{g}_{r}>0$, $\hat{l}_{r}>0$ for all the parameter values considered in this paper.
If $\delta_{\nu}^2 <0$, corresponding to linear stability, the only fixed point is $|\mathcal{A}|=0$ and the amplitude will asymptotically approach this value 
for large $x$.
In the case $\delta_{\nu}^2 >0$ we additionally find the fixed point $|\mathcal{A}|_{\mathrm{sat}}= \sqrt{\delta_{\nu}^2\,  \hat{g}_{r}/\hat{l}_{r}} \, x^{-1}  
$ and the amplitude will converge to this nonzero value for large $x$, unless the initial value 
is $|\mathcal{A}_{0}|=0$.\\ 
To solve (\ref{eq:landauamp2}), consider the factorization ansatz
 \begin{equation}
|\mathcal{A}|=A_{l}\,A_{nl}, \hspace{0.2 cm} \mathrm{with} \hspace{0.1 cm} A_{l},A_{nl}>0.
\end{equation} 
We demand $A_{l}$ to satisfy the linear equation
 \begin{equation}
\frac{\mathrm{d}A_{l}}{\mathrm{d}x} = \delta_{\nu}^{2} \hat{g}_{r}\,x^{2}A_{l}.
\end{equation} 
The corresponding solution is
 \begin{equation}\label{eq:linsol}
A_{l}=A_{l,0} \exp \Bigg\{\int_{0}^{x}\delta_{\nu}^{2} \hat{g}_{r} \,t^{2}\mathrm{d} t  \Bigg\} \equiv A_{l,0}\exp \left\{ \frac{1}{3}\delta_{\nu}^{2} 
\hat{g}_{r}\, x^{3} \right\}
\end{equation} 
with $A_{l,0}\equiv A_{l}\left(x=0\right)$. With solution (\ref{eq:linsol}), Eq. (\ref{eq:landauamp2}) can be written as
 \begin{equation}
\frac{\mathrm{d}A_{nl}}{\mathrm{d}x}=-\hat{l}_{r}x^{4}A_{l,0}^{2} \exp \left\{\frac{2}{3} \delta_{\nu}^{2}\, \hat{g}_{r} x^{3}\right\}A_{nl}^{3}.
\end{equation}  
This equation can be integrated with the result
 \begin{equation}\label{eq:nlinsol}
A_{nl}=\left[ 2 \hat{l}_{r} A_{l,0}^{2} \int_{0}^{x}   t^{4} \exp \left\{ \frac{2}{3}\delta_{\nu}^{2} \hat{g}_{r}\,  t^{3} \right\} \mathrm{d} t + 
\frac{1}{A_{nl,0}^{2}} \right]^{-1/2},
\end{equation} 
where $A_{nl,0}\equiv A_{nl}\left(x=0\right)$. Finally, with Eqs. (\ref{eq:linsol}) and (\ref{eq:nlinsol}), the final solution of (\ref{sub:landauamp}) reads
 \begin{align}\label{eq:landauampsolap}
|\mathcal{A}|\left(x\right) & =\frac{ |\mathcal{A}_{0}| \exp\left\{ \frac{1}{3} \delta_{\nu}^{2}\hat{g}_{r}\,  x^{3} \right\}}{\sqrt{ 2 \hat{l}_{r} 
|\mathcal{A}_{0}|^{2} \int_{0}^{x}  t^{4} \exp\left\{ \frac{2}{3} \delta_{\nu}^{2} \hat{g}_{r}  t^{3} \right\} \mathrm{d} t +1}}\\[0.1cm]
 & = \left[|\mathcal{A}_{0}|^{-2} \exp\left\{-\frac{2}{3}\delta_{\nu}^2 \hat{g}_{r}  x^3\right\} + \frac{\hat{l}_{r}}{\hat{g}_{r} \delta_{\nu}^2}\left(x^2 - 2 
\int_{0}^{x} t \exp\left\{\frac{2}{3}\delta_{\nu}^2 \hat{g}_{r}\left(t^3-x^3\right)\right\}\right)\right]^{-1/2}\label{eq:landauampsol2}
\end{align} 
where we defined $|\mathcal{A}_{0}| \equiv A_{l,0}\, A_{nl,0}$ and used an integration by parts in the last step.
Considering the behavior of the solutions (\ref{eq:landauampsol2}), the first aspect to notice is that the first term is only relevant for linear waves, for 
which the influence of nonlinearity, expressed through $l_{r}$ is negligible. This will be the case as long as the amplitude remains much smaller than the fixed 
point: 
$|\mathcal{A}| \ll |\mathcal{A}|_{\mathrm{sat}}$ which is the saturation amplitude of a nonlinear wave.
Thus, for small initial amplitudes and $\delta_{\nu}^2 < 0$ we obtain a linear density wave which damps exponentially due to the first term in 
(\ref{eq:landauampsol2}). 
If $\delta_{\nu}^2 >0$ (which implies viscous overstability) and $|\mathcal{A}_{0}|$ is small, the amplitude grows exponentially due to the first term as long 
as 
$|\mathcal{A}| \ll |\mathcal{A}|_{\mathrm{sat}}$. As soon as $|\mathcal{A}|$ obtains values of the order of $|\mathcal{A}|_{\mathrm{sat}}$, the second
term proportional to $l_{r}$ becomes significant and eventually damps the amplitude. For $\delta_{\nu}^2 >0$ the integral function in (\ref{eq:landauampsol2}) 
is negative and has a single minimum. The minimum marks the turnover 
of the amplitude to a power law as it asymptotically approaches the fixed point $|\mathcal{A}|_{\mathrm{sat}}$. For $x \to 0$ the integral behaves as $-x^2$ 
such that
nonlinear effects, represented by the term in round brackets, vanish.

\setcounter{figure}{0}
\setcounter{table}{0}
\setcounter{equation}{0}

\section{Pressure Tensor in the BGT86 Model}\label{sec:ptensor}
     
With \citet{bgt1986} we assume that the ring dynamics is
approximated by particles following streamlines of the form
 \begin{equation}
   \label{eq:streamline}
   r=a[1-e(a)\cos E]
\end{equation} 
in a cylindrical coordinate system $(r,\phi)$ which rotates with angular frequency $\Omega_{p}$ and origin in the planet's center of mass.
In this approximation 
 \begin{equation}
 E=m(\varphi+\Delta\left(a\right))
\end{equation} 
is the eccentric anomaly.
In the usual notation $a$ is the semi-major axis of the streamline, $e$
the eccentricity, $m$ is the azimuthal mode number, and $\Delta(a)$ is a
phase angle. 
The horizontal compression of the ring material is obtained from
 \begin{equation}\label{eq:nonlinearity}
\begin{split}
   J(\varphi,a) &\equiv \frac{\partial r}{\partial a}\\[0.1cm]
  &  \quad  = 1 - \left[e + a \frac{\mathrm{d} e}{\mathrm{d} a} \right] \cos E + m\, a\, e \frac{\mathrm{d} \Delta}{\mathrm{d} a} \sin E\\[0.1cm]
  &  \quad  \equiv 1 - q \cos E'   
\end{split}   
\end{equation} 
with the definitions
 \begin{align}
E'&=E+\gamma\\[0.1cm]
\label{eq:cosgamma}
q\cos\gamma&=\frac{\mathrm{d}(ae)}{\mathrm{d} a}\\[0.1cm]
\label{eq:singamma}
q\sin\gamma&=m\,a\,e\frac{\mathrm{d}\Delta}{\mathrm{d} a}.
\end{align} 
The variable $\gamma$ defined in this appendix should not be confused with the definition of $\gamma$ elsewhere in this paper.
The parameter $q$ is the nonlinearity parameter. For $q>1$ streamlines
start to cross.
The streamlines close in this rotating coordinate frame (in this frame the pattern
produced by the streamlines looks stationary). In an inertial frame we
have the longitude $\theta= \varphi+\Omega_p t$ and the streamlines are
described by ellipses with precessing peri-apse angle $\varpi$
 \begin{align}
   \label{eq:streamline2}
   r&=a[1-e(a)\cos \{\theta- \varpi\}]\\[0.1cm]
\theta&=\theta_{0}+\Omega t\\[0.1cm]
\varpi&= \varpi_0+\dot{\varpi}t.
\end{align} 
   Here $\Omega$ is the mean motion and $\dot{\varpi}$ is the
precession rate. Comparing the two expressions (\ref{eq:streamline}) and
(\ref{eq:streamline2}) for the orbits yields
 \begin{align}
\label{eq:LindbladOS}
m(\Omega-\Omega_p)&=\Omega-\dot{\varpi}\\
\intertext{which is the condition for a Lindblad resonance and}
\label{eq:aps0}
\varpi_0&=\theta_{0}\left(1-m\right)-m\Delta.
\end{align} 
In order to compute the components of the stress tensor, we need the
components of the radial and tangential velocities. Particles following
streamlines have the velocities (see Eq. (33) and (34) of
\citet{borderies1985})
\begin{eqnarray}
   \label{eq:kep_velocities}
   u_r&=&\Omega a e\,\sin E\\
\nonumber
u_\varphi&=&r(\Omega-\Omega_p+2\Omega e \,\cos E).
\end{eqnarray}
The hydrodynamic pressure tensor is defined as 
 \begin{equation}
 \label{eq:pressure_tensor}
 P_{\alpha\beta}=p\delta_{\alpha\beta}-2\eta
D_{\alpha\beta}-\delta_{\alpha\beta}\zeta\vec\nabla\cdot\vec u\,,
\end{equation} 
with the trace--less shear tensor
 \begin{equation}
 \label{eq:shear_tensor}
 D_{\alpha\beta} =
\frac{1}{2}\left[\partial_{x_{\beta}} u_{\alpha}+\partial_{x_{\alpha}} u_{\beta}-\frac{2}{3}\delta_{\alpha\beta}\vec\nabla\cdot\vec
u \right].
\end{equation} 
In these expressions $p$ is the isotropic pressure and $\eta$ and $\zeta$ are the dynamic shear and bulk viscosities, respectively. Further, 
$\delta_{\alpha\beta}$ is the Kronecker symbol
and $\vec u$ denotes the velocity in the ring plane.
The components of the pressure tensor we need are
\begin{eqnarray}
   \label{eq:shear_tensor_compontents}
P_{rr}&=&p+\left(\frac{2}{3}\eta-\zeta\right)\vec\nabla\cdot\vec u
-2\eta\partial_{r} {u_r}\\[0.1cm]
P_{r\varphi}&=&-2\eta D_{r\varphi}
\end{eqnarray}
with
 \begin{equation}
D_{r\varphi}=\frac{1}{2}\left[
\frac{1}{r}\partial_{\varphi} u_{r}+ \partial_{r} u_{\varphi}-\frac{u_{\varphi}}{r}\right].
\end{equation} 
From Eqs. (\ref{eq:kep_velocities}) and (\ref{eq:streamline}) we find
\begin{eqnarray}
   \label{eq:velocity_derivatives}
   \frac{1}{r}\partial_{\varphi}u_r&=&m\Omega e \cos E=O(\Omega e, e^2)\\[0.1cm]
\partial_{r}u_{\varphi}&=&\left[\frac{\Omega}{2J}-(\Omega+\Omega_p)\right]+O(e^2,\Omega
e)\\[0.1cm]
\frac{u_\varphi}{r}&=&(\Omega-\Omega_p)+O(\Omega e, e^2)\,,
\end{eqnarray}
where $\partial_r\equiv\frac{1}{J}\partial_a$ was used.
Thus we have
 \begin{align}
   \label{eq:drphi_sol}
   D_{r\varphi} &= \frac{\Omega}{4J}(1-4J),\\[0.1cm]
   \label{eq:nabla_u}
   \vec\nabla\cdot\vec u &\equiv
\frac{1}{r}\partial_{r}\left(ru_r\right)+\frac{1}{r}\partial_{\varphi} u_{\varphi}\\[0.1cm]
&=\frac{\Omega q}{J}\sin E'+O(\Omega e, e^2).
\end{align} 
Therefore
\begin{eqnarray}
   \label{eq:shear_tensor_compontents_1}
   P_{rr}&=&p-\left(\frac{4}{3}\eta+\zeta\right)\frac{\Omega q}{J}\sin
E'+O(\Omega e, e^2)\\[0.1cm]
P_{r\varphi}&=&-\eta\frac{\Omega}{2J}(1-4J)+O(\Omega e, e^2)\label{eq:shear_tensor_compontents_2}.
\end{eqnarray}
We define
\begin{equation}\label{eq:sigrel}
 \sigma(r)=\frac{\sigma_{0}(r)}{J} 
\end{equation}
where $\sigma(r)$ (in the following $\sigma$) denotes the perturbed surface density as it has been used elsewhere in this paper. 
The quantity $\sigma_{0}(r)$ 
is the radially dependent (background) surface density which is required to maintain the viscous angular momentum luminosity in the wave zone $r>r_{L}$ which 
differs from its value inside the resonance ($r<r_{L}$) due to the satellite torque which is excited 
at the resonance and the presence of the density wave [see Eqs. (24), (25) and (29) in BGT86]. 
The value of $\sigma_{0}(r)$ inside the resonance which equals the value of the unperturbed ring is denoted by $\sigma_{0}^{-}$.
We apply definition (\ref{eq:shearvis}) for viscosity and the isothermal ideal gas relation for pressure such that
 \begin{equation}
\eta= \sigma  \nu_{0}\left(\frac{\sigma}{\sigma_{0}^{-}}   \right)^\beta
\end{equation} 
and 
 \begin{equation}
 \begin{split}\label{eq:ceff}
p  & =  \left[\frac{\partial p}{\partial \sigma}\right]_{0}  \sigma\\
  \quad  & =   p_{\sigma} \, \sigma
\end{split}
\end{equation} 
where $p_{\sigma}$ assumes the values given in Table \ref{tab:scalings}. 
Inserting (\ref{eq:sigrel})-(\ref{eq:ceff}) in the pressure tensor components (\ref{eq:shear_tensor_compontents_1}) and (\ref{eq:shear_tensor_compontents_2}) 
yields
\begin{subequations}
 \begin{align}
   \label{eq:shear_tensor_compontents_3}
   P_{rr}&= \sigma p_{\sigma}  -  \sigma \nu_{0} \left(\frac{\sigma}{\sigma_{0}^{-} }\right)^{\beta}  \left(\frac{4}{3}+\gamma \right) \frac{\Omega 	q}{J} 
\sin E',  \\[0.1cm]
P_{r\varphi}&= -  \sigma\nu_{0} \left(\frac{\sigma}{\sigma_{0}^{-} }\right)^{\beta} \frac{\Omega}{2 J}\left(1-4 J \right).
\end{align} 
\end{subequations}
The viscous coefficients needed for the computation of density wave profiles, according to the method described in Section IVa) in BGT86, are [Equations (17), 
(18) and (25) in BGT86]\footnote{Note that the quantity $\Sigma_{0}(r)$ in BGT86 corresponds to our quantity $\sigma_{0}(r)$. The value $\sigma_{0}^{-}$ in our 
paper corresponds to $\Sigma_{-}$ in BGT86.}
 \begin{align}
   \label{eq:viscous_coefficient_1}
\mathcal{T}_1 & \equiv\frac{1}{2\pi \sigma_{0}(r)} \int_0^{2\pi} \mbox{d}E'\left[P_{rr}\sin,
E'+2P_{r\varphi}\cos E'\right]\\
\mathcal{T}_2 & \equiv\frac{1}{2\pi \sigma_{0}(r)}\int_0^{2\pi}\mbox{d}E'\left[-P_{rr}\cos,
E'+2P_{r\varphi}\sin E'\right]\label{eq:viscous_coefficient_2}\\
a_{r \varphi} & \equiv\frac{1}{2\pi \sigma_{0}(r)}\int_0^{2\pi}\mbox{d}E'P_{r\varphi}.\label{eq:viscous_coefficient_3}
\end{align} 

The behavior for \emph{small values of $q$} can be derived if the above integrals are evaluated with expressions for $P_{rr}$ and $P_{r\varphi}$ that are 
linearized in $q$.
To this extent we use the expressions 
 \begin{align}\label{eq:nuexp}
\eta & = \nu_{0} \,\sigma_{0} \left[ \frac{\beta+1}{J} -\beta \right]\\
p & = \frac{p_{\sigma} \,\sigma_{0}}{J}\label{eq:pexp},
 \end{align}
for viscosity and pressure which are valid for small $q$. Note that in the limit $q\to 0$ which occurs for $r\to r_{L}$ we have $\sigma_{0}(r)=\sigma_{0}^{-}$.
Inserting (\ref{eq:nuexp}), (\ref{eq:pexp}) in the pressure tensor components (\ref{eq:shear_tensor_compontents_1}), (\ref{eq:shear_tensor_compontents_2}) and 
linearizing in $q$ yields
\begin{align}
P_{rr} & = p_{\sigma} \, \sigma_{0} + q \left[ p_{\sigma} \sigma_{0} \cos E' - \left(\frac{4}{3} + \gamma \right) \nu_{0} \,\Omega\, \sigma_{0} \sin E'\right]\\
P_{r\varphi} & = \frac{1}{2} \nu_{0} \,\Omega\, \sigma_{0} \left[3 + q \left(2+3\beta\right) \cos E' \right]
\end{align}
The integrals (\ref{eq:viscous_coefficient_1})-(\ref{eq:viscous_coefficient_3}) then yield in \emph{scaled form} (scalings from Table \ref{tab:scalings})
\begin{subequations}
 \begin{align}
  \mathcal{T}_1 & = \frac{3}{2}\left[\beta - \frac{1}{3}\left(\gamma-\frac{2}{3}\right)\right]\nu_{0} \, q\label{sub:t1lin}\\ 
  \mathcal{T}_2 & = -\frac{1}{2}p_{\sigma} \, q\\ 
  a_{r \varphi} & = \frac{3}{2}\nu_{0}
 \end{align}
\end{subequations}
As discussed in Section \ref{sec:bgtcomp} linear instability of a density wave occurs if $\mathcal{T}_1>0$ for small $q$.
From (\ref{sub:t1lin}) we find that the criterion for instability equals the hydrodynamic for viscous overstability (\ref{eq:marstabnlo}) [\citet{schmit1995}], 
in agreement with our model.

\end{document}